\documentclass[12pt,preprint]{aastex}%
\usepackage{lscape}

\begin{document}

\title{Stellar Orbits Around the Galactic Center Black Hole}

\author{A. M. Ghez\altaffilmark{1}, S. Salim, S. D. Hornstein, A. Tanner, J. R. Lu,
M. Morris, E. E. Becklin, G. Duch\^ene}

\affil{UCLA Division of Astronomy and Astrophysics, Los Angeles, CA 90095-1574}
\email{ghez, samir, seth, tanner, jlu, morris, becklin, duchene@astro.ucla.edu}
\altaffiltext{1}{Institute of Geophysics and Planetary Physics, University of 
California, Los Angeles, CA 90095-1565}

\begin{abstract}

We present new diffraction-limited images of the Galactic Center, obtained
with the W. M. Keck I 10-meter telescope.  Within 0\farcs 4 
of the Galaxy's central dark mass, 17 proper motion stars, 
with $K$ magnitudes ranging from 14.0 to 16.8, are identified 
and 10 of these are new detections (6 were also independently discovered
by Sch\"odel et al.\ 2003).  In this sample, 
three newly identified (S0-16, S0-19, and S0-20) and four previously known
(S0-1, S0-2, S0-4, and S0-5) sources have
measured proper motions that reveal orbital solutions.

Orbits are derived {\it simultaneously} so that they jointly constrain 
the central dark object's properties: its mass, its position,
and, for the first time using orbits, its motion on the plane of the sky.
This analysis pinpoints the Galaxy's central dark mass 
to within 1.3 mas (10 AU) and limits its proper motion to 
1.5 $\pm$ 0.5 mas y$^{-1}$ (or equivalently 60 $\pm$ 20 km s$^{-1}$) 
with respect to the central stellar cluster.
This localization of the central dark mass is consistent with our
derivation of the position of the radio source Sgr A* in the infrared
 reference frame 
($\pm$ 10 mas), but with an uncertainty that is a factor 8 times smaller, which
greatly facilitates searches for
near-infrared counterparts to the central
black hole.  Consequently, one previous claim for such a counterpart can now be
ascribed to a close stellar passage in 1996.  Furthermore, 
we can place a conservative upper limit of 15.5 mag
on any steady-state counter-part emission.
The estimated central dark mass from orbital motions is 
$3.7 (\pm 0.2 ) \times 10^6 (\frac{R_o}{8kpc})^3 M_{\odot}$; this is a more
direct measure of mass
than those obtained from velocity dispersion measurements, which are as much
as a factor of two smaller.  
The Galactic Center's distance,
which adds an additional 19\% uncertainty in the estimated mass, is now  
the limiting source of uncertainty in the absolute mass.
For stars in this sample, the closest approach 
is achieved by S0-16, which came within a mere 
45 AU (= 0.0002 pc = 600 R$_{s}$) at a velocity 
of 12,000 km s$^{-1}$.  This increases the inferred dark mass density by
four orders of magnitude compared to earlier analyses based on velocity and acceleration vectors, making the Milky Way the strongest existing case for a supermassive black hole at the center of a normal type galaxy.

Well determined orbital parameters for these seven Sgr A* cluster stars 
provide new constraints on how these apparently massive, young ($<$10 Myr) 
stars formed in 
a region that seems to be hostile to star formation.
Unlike the more distant He-I emission line stars - another 
population of young stars in the Galactic Center - that appear
to have co-planar orbits, the Sgr A* cluster stars have orbital
properties (eccentricities, angular momentum vectors, and apoapse
directions) that are consistent with an isotropic distribution.
Therefore many of the mechanisms proposed for the formation of
the He-I stars, such as formation from a pre-existing disk, are
unlikely solutions for the Sgr A* cluster stars.
Unfortunately, alternative theories for producing young stars, or 
old stars that look young, in close proximity to a central supermassive 
black hole, are all also somewhat problematic.
Understanding the apparent
youth of stars in the Sgr A* cluster, as well as the more distant He~I
emission line stars, has now become one of the major outstanding issues in
the study of the Galactic Center.

\end{abstract}


\pagebreak

\section{Introduction}

The proximity of our Galaxy's center (8 kpc, Reid 1993) presents an 
opportunity to build a case for a supermassive black hole and to 
study the black hole's environment and its effects thereon with much higher 
spatial resolution than can be brought to bear on any other galaxy.
The first hint of a central concentration of dark matter in the Milky Way came 
from radial velocity measurements of ionized gas located in a three-armed 
structure known as the mini-spiral, which extends from the center
out to $\sim$1-3 pc (Lacy et al.\ 1980).   Concerns that the measured
gas motions were not tracing the gravitational potential were quickly 
allayed by radial velocity measurements of stars, which are not susceptible 
to non-gravitational forces (McGinn et al.\  1989; Sellgren et al. 1990; 
Haller et al.\ 1996; Genzel 
et al.\ 1997).  These early, low angular resolution, dynamical measurements of 
the gas and stars at the center of the Milky Way suggested the presence
of $\sim 3 \times 10^6 M_{\odot}$ of dark matter and confined it to within a radius
of $\sim$0.1 pc.  The implied minimum dark matter density of 
$\sim 3 \times 10^{9} M_ {\odot}  pc^{-3}$, however, 
still allowed a cluster of dark objects, such as neutron stars
or stellar mass black holes, as 
one of the alternatives to a single supermassive black hole, 
because the measurements did not force the cluster's lifetime to be 
shorter than the age of the Galaxy (Maoz et al.\ 1998).

Significant progress has been made recently with
diffraction-limited near-infrared studies of the central stellar 
cluster. The first phase of these experiments yielded proper motion 
velocities
(Eckart \& Genzel 1997; Ghez et al.\ 1998),
which suggested that $2.6 (\pm 0.6) \times 10^6 M_{\odot}$ of dark matter is 
confined to within 0.015 pc.
This increased
the implied minimum dark matter density by 3 orders of magnitude to
$10^{12} M_{\odot} pc^{-3}$ and eliminated a cluster of dark objects
as a possible explanation of the Galaxy's central dark mass
concentration (Maoz et al.\  1998), but still left 
the fermion ball hypothesis (e.g., Tsiklauri \& Viollier 1998; Munyaneza \&
Viollier 2002) as an alternative to a single supermassive black hole.
The velocity dispersion measurements also localized the dark
matter's centroid to within 100 mas and at a position 
consistent with the nominal
location of the unusual radio source Sgr A* (Ghez et al.\ 1998), whose
emission is posited to arise from accretion onto a central supermassive black
hole (e.g., Lo et al.\ 1985).
The detection of acceleration for three stars -- S0-1, S0-2, and S0-4 --
localized the dark mass to within 30 mas, 
increased the dark matter's minimum density to $10^{13} M_{\odot} pc^{-3}$,
and thereby further strengthened both the case for a supermassive black hole and
its association with Sgr A* (Ghez et al.\ 2000; Eckart et al.\ 2002).

Deviations from linear motions also initiated a new phase
for these proper motion experiments, that of direct orbital studies.
By making a number of
assumptions, including fixing the central mass
to the value obtained from the velocity dispersion analysis and its
location to that inferred for Sgr A* by Menten et al.\ (1997),
Ghez et al.\ (2000) and Eckart et al.\ (2002) obtained the first crude
orbital solutions; these experiments revealed that orbital periods for
S0-2 and S0-1 could be as short as 15 and 35 years, respectively.
With a larger fraction of the orbit being traced, more precise orbital analyses have been 
carried out
for S0-2 by Sch\"odel et al.\ (2002), who dropped the mass assumption, and by
Ghez et al.\ (2003), who dropped both the mass and center of
attraction assumptions and added radial velocity measurements.
These orbital solutions suggested that S0-2 made a closest approach of 
0.0006 pc (120 AU) in 2002 and that its orbit encloses a central mass of 
$3.7 (\pm 1.5) \times 10^6 M_{\odot}$ (Sch\"odel et al.\ 2002) or
$4.0 (\pm 0.6) \times 10^6 M_{\odot}$ (Ghez et al.\ 2003)\footnote{
The uncertainties in the estimated mass in Ghez et al.\ (2003) are 
a factor of 2.5 smaller than that in Sch\"odel et al.\ (2002),
despite the two additional free model parameters introduced by fitting
for the center of attraction and the shorter time baseline.  This   
is primarily due to the higher astrometric accuracy of the Keck data set,
rather than the inclusion of radial velocity measurements.}, 
which is somewhat 
higher than that found from the velocity dispersion measurements.
Possible causes of this discrepancy include (1) inaccuracies in the 
assumptions made in the use of the 
velocity-dispersion-based projected mass estimators
about the 
stellar cluster's number density distribution and/or the level of anisotropy
(Genzel et al.\ 2000; Figer et al. 2003),
(2) inaccuracies in the orbital fit assumptions, such as 
the central mass distribution being point-like and 
at rest with respect to the central stellar cluster, and (3) systematic 
errors in either the overall velocity dispersion measurement for the central 
stellar cluster or S0-2's individual positional measurements.  
Further 
measurements are necessary to determine whether this 
2 $\sigma$ difference 
in the estimates of the central dark mass produced by
the two methods is real and, if so, what its origin is.

While the detection of spectral lines in S0-2 provided full dynamical 
information,
it also offered insight, for the first time, 
into the nature of this star that is orbiting in
such close proximity to the central dark mass.
S0-2's spectral features are consistent with those of an O8-B0 dwarf, suggesting
that it is a massive ($\sim$15 $M_{\odot}$), young ($<$10 Myr), main sequence
star (Ghez et al.\ 2003).  Less direct measurements of other stars within the 
central 1$\tt''$ $\times$ 1$\tt''$, which are known collectively
as the Sgr A* stellar cluster, imply that these stars might be
similarly young; 
specifically, their similar 2 $\mu$m luminosities and the 
lack of CO absorption in spectra of individual stars (Genzel et al.\ 1997; 
Gezari et al.\ 2002) or in integrated spectra of the Sgr A* stellar cluster 
(Eckart, Ott, \& Genzel 1999; Figer et al.\ 2000) lead to the conclusion
that they, like S0-2, 
have hot photospheres consistent with massive young stars. 

While the presence of young stars in close proximity to our Galaxy's
supermassive black hole 
has long been recognized as a problem in the context of the young He~I 
emission-line stars (Sanders 1992; Morris 1993), this problem is much
worse for the Sgr A* cluster stars, whose
distances from the black hole are an order of magnitude smaller.
At S0-2's apoapse distance of 0.01 pc, inferred from the orbital 
solutions,
the Roche density is $10^{14}$ cm$^{-3}$,
whereas the maximum density determined for even the nearby
circumnuclear disk, located at radii of $\sim$ 1-3 pc, is only about 10$^5$ -
10$^6$
cm$^{-3}$ (e.g., Jackson et al.\ 1993; Christopher \& Scoville 2003).
Furthermore, at present, the region over which S0-2
is currently orbiting contains only a very low-density
plasma, as evidenced by weak Br-$\gamma$ line emission (Figer et al.\ 2000;
Gezari et al.\ 2002).
Several ideas proposed to account for the apparently young 
He~I emission-line stars may be applicable to the Sgr A* cluster stars and
they fall into the following three broad categories: 
(1) the stars are indeed young and formed in-situ,
which requires much higher local gas densities in the recent 
past in order to enable star formation to proceed in the black hole's strong tidal field
(e.g., Levin \& Beloborodov 2003),
(2) the stars are young and formed at larger radii, where the black hole's
tidal effects are small, and underwent rapid orbital migration inwards
(e.g. Gerhard 2001; Kim \& Morris 2003; Hansen \& Miloslavjevic 2003),
and (3) the stars are old (initially formed long ago), but their
appearance has been altered, due to interactions with the 
local environment, such that they appear young but 
have had sufficient time to migrate inwards from their original
birth place (e.g., Morris 1993; Genzel
et al.\ 2003a).  
%of ``nature", in which case
%star formation did indeed occur very recently, and ``nurture",
%in which case the extreme environment at the Galactic center has
%significantly altered the evolution of these stars. 
Stellar kinematics produced by these mechanisms are likely to differ. 
Well constrained orbits for a set of stars in the Sgr A* cluster
would allow a direct examination of the cluster's kinematics and therefore 
would provide important insight into how these stars formed and came to be on
their present orbits.

This paper reports new proper motion measurements obtained
with the W. M. Keck 10-meter telescope for 4 previously known stars 
(S0-1, S0-2, S0-4, S0-5) and for 3 newly identified stars
(S0-16, S0-19, S0-20).   The trajectories of all of these stars show 
significant 
curvature or linear acceleration, thus allowing 
the first {\it simultaneous} orbital analysis for multiple stars making their
closest approaches to the central dark mass.
Section 2 describes the observations, which now cover an 8-year time baseline.
Section 3 provides the details and results for source identification, 
astrometry,
and the orbital fits, which, for the first time, allow for the dark mass'
motion on the plane of the sky.  Section 4.1 discusses the
constraints that the orbital parameters offer on the nature of the central 
dark mass distribution, which has become the best case yet for a supermassive
black hole at the center of any normal type galaxy and whose mass, position 
and motion in the infrared reference frame are determined with 
unprecedented accuracy.  Finally, Section 4.2 explores how the direct 
measurements of orbital dynamics impact the question of the 
origin of the central
stellar cluster. 

%*** in this last paragraph of the intro, you might whet the appetite of
%the reader
%*** a bit more by stating that we can now nail down the position and
%strongly
%*** constrain the motion of the central mass concentration.   Also, we
%are filling
%*** out the picture of the dynamics of the apparently young stars in
%the central parsec,
%*** in terms of their isotropy and eccentricity distribution, and thus
%we can shed
%*** some light on their origin.

\section{Observations }

New $K$[2.2 $\mu m$]-band speckle imaging observations of the Galaxy's 
central stellar cluster were obtained with the W. M Keck I 10-meter telescope 
using the facility near-infrared camera, NIRC (Matthews \& Soifer 1994; 
Matthews et al.\ 1996) 
%during 9 separate observing runs between 2000 April
%and 2003 April (see Table~\ref{tab_obs}).
on the nights 
of 2000 April 21, 2000 May 19-20, 2000 July 19-20, 2000 Oct 18, 
2001 May 7-9, 2001 July 28-29,
2002 April 23-24, 2002 May 23-24 \& 28-29, 2002 June 2, 2002 July 19-20,
2003 April 21-22, 2003 July 22-23, and 2003 September 7-8.
These data sets were collected and analyzed similarly to the data sets
obtained between 1995 and 1999 for this project
(see Ghez et al.\ 1998, 2000 for details).  In summary, short ($t_{exp}$ = 
0.1 sec) 
exposures were obtained in sets of $\sim$200, resulting in a total of 
$\sim$7,000 exposures per observing run.  
Each frame, with a scale of 20.396 $\pm$ 0.042 mas pixel$^{-1}$ (see Appendix B) and 
a corresponding 
field of view of 5.$\tt''$22 x 5.$\tt''$22, was sky-subtracted, flat fielded,
bad-pixel-corrected, corrected for distortion effects, and magnified by a 
factor of two.  In sets of 200, the frames were shifted to the location of the 
brightest
speckle of IRS 16C (K = 9.8 mag) and combined to create intermediate shift-and-add (SAA) maps,
which have point spread functions (PSF) that can be described as containing 
a diffraction-limited core on top of a seeing halo. 
These were then combined after applying a seeing cut, which required that the
seeing halo FWHM be less than $\sim$-0.$\tt''$4 - 0.$\tt''$6, depending on
the overall quality of the night.  Final SAA maps have point
spread functions composed of a diffraction-limited core ($\theta$ $\sim$ 0.$\tt''$05),
containing $\sim$ 4\% of the radiation, on top of a halo that has a FWHM of 
$\sim$0.$\tt''$4.
In addition to averaging all the data from each run 
to produce a final SAA map, these data were divided into three sub-sets to 
construct ``sub-maps", which were used to determine positional and brightness 
uncertainties.

\section{Data Analysis \& Results}

\subsection{Source Identification}

\begin{figure}
\epsscale{1.0}
%\plotone{figures/mag00may_corr_label.eps}
\plotone{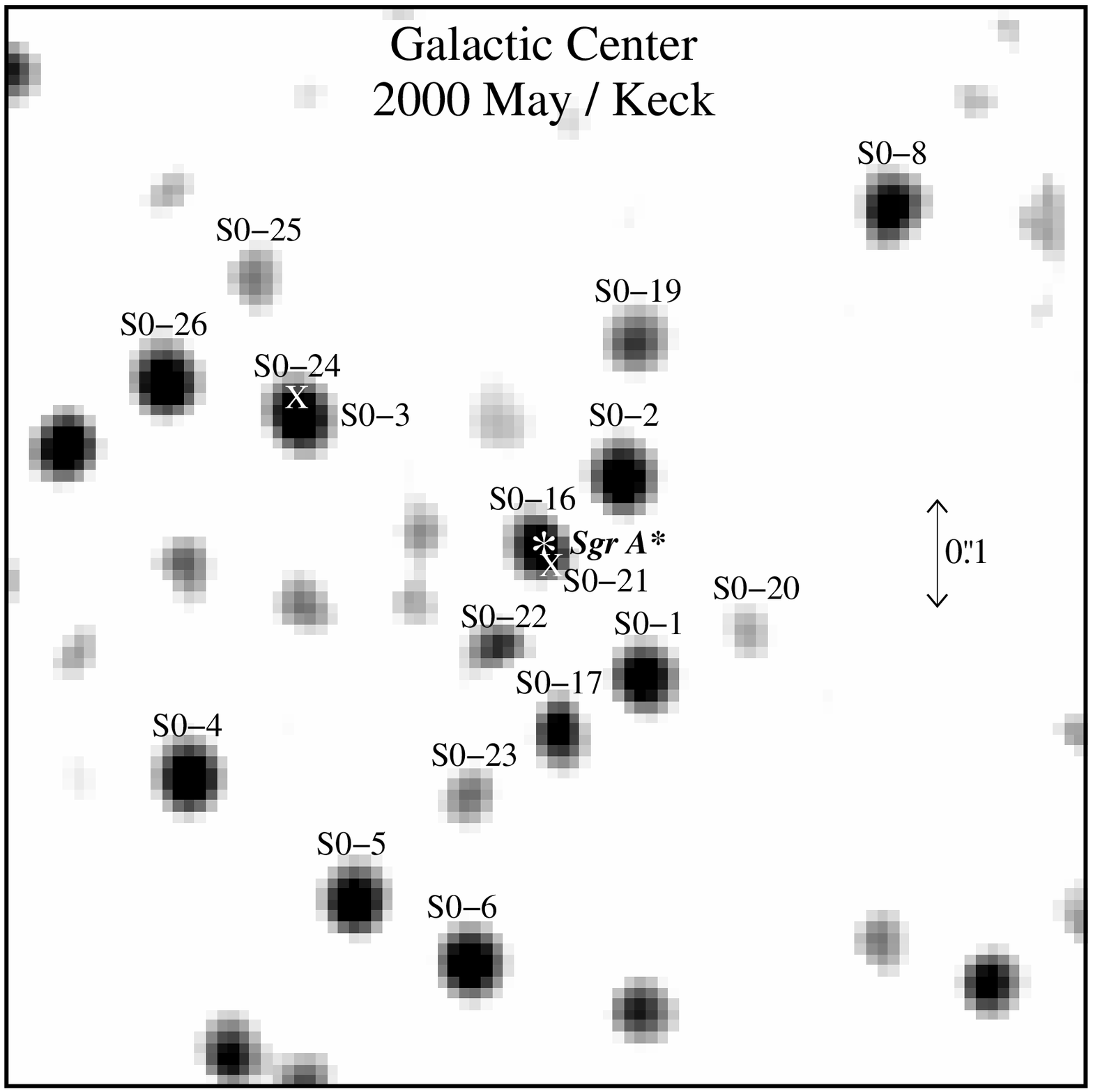}
\figcaption{ 
The central 1$\tt''$ $\times$ 1$\tt''$ of the cross-correlation 
(or match filter) map for the 2000 May data set.  Of the 17 sources
identified in this study by the criteria described in \S3.1, 15
are seen in this map.  The remaining 2, marked with crosses,
are missed in this particular map due to confusion with 
a brighter nearby source.  
An asterisk denotes the black hole's dynamically determined position 
(see \S3).
The criteria used for source identification are still quite
conservative as there are several unlabeled peaks that appear to be
real sources, within 0\farcs 4 of Sgr A*.
\label{gc_image}}
\end{figure}

Sources are identified using the same procedure described by Ghez et 
al. (1998), with a few minor modifications.  
As in Ghez et al.\ (1998), a ``match
filter" is applied to each image, by cross-correlating the image with
the core of its PSF, out to a radius of 0.$\tt''$06 
(see Figure~\ref{gc_image}).  
In a first pass at source identification,
correlation
peaks larger than a threshold value are flagged as stars.
Once
stars are identified, a second lower threshold value is used to track 
these stars in images in which they were not identified with the first
threshold value; this second pass search is limited to within a specified
radius of the predicted position.  Positions of sources found in either
the first or second pass search are 
estimated based on the correlation map peak and only sources that are 
identified
in at least 3 epochs are included in our final proper motion sample.
While in Ghez et al.\ (1998)
the predicted position for the second pass source search was simply the position found in the first
pass, here we use any kinematic information 
available from the first pass to define this predicted position.
Two other modifications change only the values used in the
algorithm.  
We lowered the first pass threshold correlation value for 
source identification from 0.7 to 0.5, which allows fainter sources to
be identified, and we have decreased the second pass search
area radius from 0\farcs 07 to 
the uncertainty in the predicted position 
(with the constraint that it must be at least 0\farcs 01 and no more than
0\farcs 07),
due to the increased number of sources that are being tracked.
Positions are now estimated using gaussian fits, 
as opposed to a simple centroiding algorithm.
The final modification requires that each source be detected in all three
sub-maps (see \S2); first and second pass sources had sub-map 
correlation thresholds of 0.3 and 0.2, respectively.

Photometric values are estimated using two methods.  First, simple
aperture photometry, as described in Ghez et al.\ (1998), 
is applied
to help track the sources through the data set.  Second, PSF fitting 
with StarFinder (Diolaiti et al.\ 2000) is implemented
and these are the values (average and rms) reported in Table~\ref{tab_sum}.  
StarFinder and aperture photometry produce
the same results for bright (K $\lesssim$ 15.2) sources, but,
for the fainter sources, the StarFinder results are 
somewhat fainter ($\sim$ 0.2 mag) and more precise, 
due to the stellar confusion.
In this study, the zero points are established based on 
Blum et al.\ (1996) measurements of 
IRS 16C (K=9.83 mag), IRS 16NW (K=10.03 mag), and IRS 16NE (K=9.00 mag), 
which results in magnitudes that 
are $\sim$ 0.2 mag fainter than than those reported in Ghez et al.\ 1998,
which relied on IRS16NE only for a zero point estimate.

While many sources are identified and tracked over our 
entire $\sim$5$\tt'' \times$5$\tt''$ field of view, this study is limited to 
sources within a radius of 0\farcs 4 of the infrared position for Sgr A* 
(see Appendix B); the radius is set by the
criterion that all stars with accelerations of 2 mas y$^{-2}$ or greater
should reside within this region, 
assuming a mass, $M$, of $3.7 \times 10^6 M_{\odot}$ (see \S4.1),
 or equivalently  $r_{max}^2 = G M / a_{min}$.

This procedure identifies 17 proper motion sources ($ K \lesssim 16.8 $), of which 10 
are newly discovered in this 
study\footnote{We note that after this paper was submitted for 
publication,
6 of the 10 new sources were also reported by Sch\"odel et al.\ (2003). } 
and all of which are shown
in Figure 1.
%It is worth noting that more sources are detected here within a radius of 
% 0\farcs 4
%than in recent VLT maps, in which 14 sources with K$<$17 mag and 19
%sources with H$<$19.4 mag are detected over the same radii.
The new sources are fainter than the sources in this study that were 
previously published ($K_{new} \gtrsim 15.1$ mag),  
with only one exception (S0-8, which is located at the largest projected
separation).
Among the original 
proper motion sample reported in Ghez et al.\ (1998), there are many other 
sources comparably faint to the newly discovered proper motion sources, but at 
larger radii; the reason for this is that at the center
of the maps source confusion lowers the correlation values and reduces the 
sensitivity to faint sources using our source identification technique.
The new source detections are therefore a consequence
of our lower correlation thresholds and, as can be 
seen in Figure 1, these thresholds 
are still fairly conservative, since a number of additional 
sources are seen in the cross-correlation maps.  
While the previously known sources are detected
in all the maps, the new sources are not, 
due to the variation in the maps' sensitivities and, occasionally,
confusion with a brighter source (see footnotes in Table~\ref{tab_sum}).
Nonetheless, sources as faint as $\sim$ 16.7 are detected in
the majority of maps in the second pass for source detections and
sources brighter than $\sim$ 15.5 mag are detected in all
maps in the first pass. 
%Reliably identifying and tracking stars 
%fainter than $K \sim$15 mag through such a crowded region requires multiple 
%maps per year, which we began to obtain in 1998.
Table~\ref{tab_sum} lists the properties of all the 
detected sources in our sample; the new sources are named
according to the convention introduced in Ghez et al.\ 1998 and summarized
in Appendix A.

\subsection{Astrometry}

Stellar astrometry is derived in three separate steps.  
First, centroid positions on the
correlation peaks provide estimates of the stars' locations in each of the
maps.  
Uncertainties in these locations are estimated 
based on the rms of their locations in the
3 sub-maps created for each map (see \S2) and have
an unweighted average value of
$\sim$2 mas for the brightest stars (K $<$ 15 mag) and 
$\sim$5 mas for the K$\sim$ 15.5 - 15.7 mag stars.
Second, the coordinate system
for each map is transformed, with the application of a net translation
and rotation, to a common local reference frame.  As in Ghez et al.\ (1998), 
this transformation is determined by minimizing the net
motion of the measured stars, but with the three modifications:
(1) rather than using all stars detected in the central 
$\sim$5$\tt'' \times$5$\tt''$, we now exclude those stars
within 0\farcs 5 of Sgr A* as well as those that have correlation 
values less than 0.7, (2) the common local reference
frame is
now chosen to be the map obtained at the middle epoch, 1999 July, instead of 
the 1995 June map, in order to minimize the uncertainty in the coordinate
transformations, which increase with the temporal distance from the reference
epoch,
and (3) a second pass through the minimization
process is applied using initial estimates of the proper motions 
from the first pass.   
As described in Ghez et al.\ (2000), 
positional uncertainties associated with this transformation of
the relative positions of stars into a common coordinate system
are estimated by a half-sample bootstrap method.
These uncertainties are minimized
at the center of the field of view, which is where the stars reported here 
were 
always observed, and that decrease with both the number of stars included 
in estimating the transformation (48 - 104 stars) and 
the closer in time the epoch is to 
the reference epoch.
The values of these uncertainties
for the stars in this study 
range from zero, for the reference epoch of 1999 July, to
0.44, for epochs close to 1999 July, and up to 2.5 mas, for
data sets at the extrema in time of our experiment.  
Compared to the centroiding uncertainties, the transformation uncertainties 
are negligible for the faintest stars and are, in some epochs, 
comparable to that of the brightest stars in our sample.  The 
uncertainties from the
centroiding and transformation processes are added in quadrature to produce the
final relative position uncertainties.  It is these relative positions 
that are used in the orbital analysis presented in \S3.3.1 and their weighted
averages are presented in column 10 of Table~\ref{tab_sum}.  The third
and final aspect of the astrometric measurements is transforming 
the relative positions and orbital solutions from 
the 1999 July map's coordinate system to an absolute coordinate system 
using measurements of sources with known absolute astrometry, as 
described in Appendix B; this final transformation has been applied
to the orbital solutions presented in \S3.3.2.

Stellar confusion not only prevents faint sources from being detected (as
discussed above), but can also generate astrometric biases.
Fortunately, the sources' high proper motions easily reveal the 
underlying biasing sources at later or earlier times, so that these
biases can be recognized.  We exclude the biased
points in a three step procedure.  First, 
measurements of stars during epochs in which they are 
obscuring other stars are temporarily excluded; at this point, stars with 
remaining measurements in less than three years (S0-3 and S0-21) are 
removed from further dynamical analysis due to our inability to assess
the possible effects of astrometric biases.
% # of years with no confusion
% S0-3	0 years
% S0-21 only detected in 2 years
% S0-22 only detected in 3 years
% S0-17 4 years ('96, '99, '00, '01)
Second, with this vetted data set, we carry out preliminary linear or orbit model fits,
in which no parameters are constrained (see \S3.3.1).
Third, we return to the original data set and remove 
those points that are offset by more than 3$\sigma$ from the preliminary 
best fit model.
Column 5 of Table~\ref{tab_sum} summarizes the number of remaining points that 
are used in the fits described in \S3.3.2.

\subsection{Orbital Analysis}

Significant curvature or linear acceleration in the plane of the sky 
is detected for 7 of the 17 sources listed in Table~\ref{tab_sum},
using the criterion defined by Ghez et al.\ (2000), 
but accounting for the differing number of epochs by considering
$\chi^2_{tot}$ instead of $\chi^2_{dof}$ (see column 11 in 
Table~\ref{tab_sum}).  
Stars are considered to show significant deviations from linear 
proper motion, if they have $\Delta \chi^2_{tot}$, which is 
the difference between the total $\chi^2$ value
resulting from the best linear fit and the total $\chi^2$ from the
best second order polynomial fit, greater than 15.
S0-1, S0-2, S0-4, S0-5, S0-16, S0-19, and, S0-20, satisfy this
criterion and 
have accelerations in the plane of the sky 
of at least 2 mas y$^{-2}$ and as much as 1500 mas y$^{-2}$
and, of these, S0-2, S0-16, and S0-19 have measurable higher order 
positional derivatives. 
We therefore carry out 
orbital fits for these 7 stars with a model described in \S3.3.1 and 
with resulting
orbital parameters given in \S3.3.2.

\subsubsection{Model \& Method}

We assume a model in which the gravitational potential arises from a 
single dominant point mass, 
which allows multiple stars to contribute
simultaneously to the solution for the following properties of the 
central object: 
\begin{itemize}

\item Mass ($M$)
\item Location ($r_{RA}$, $r_{DEC}$) 
\item Linear motion on the plane of the sky ($v_{RA}$, $v_{DEC}$).  

\end{itemize} 
In this analysis, the point source's distance ($R_o$) and its linear motion 
along the line of sight ($v_z$) are not solved for; to get $M$,
$R_o$ is assumed to be 8 kpc (Reid 1993), while $v_z$ is set to 0.
Setting v$_z$ equal to 0 km s$^{-1}$ is reasonable given
that the resulting limits on the values of v$_{RA}$ and v$_{DEC}$ 
($\lesssim$76 km s$^{-1}$) are
comparable to the uncertainties on the radial velocity measurements for S0-2
($\sim$40 km s$^{-1}$, Ghez et al.\ 2003), which
are the only radial velocities used in this analysis;
furthermore, Figer et al.\ (2003) find an average v$_z$ for a set of
 cool stars at the Galactic Center consistent with 0 to within 11 km s$^{-1}$. 
In addition to these 5 common free parameters, there are the following
6 free parameters for each star: 
\begin{itemize}
\item Period ($P$), which, when 
combined with the estimate of the central dark mass, yields the angular
semi-major axis ($A$) 
\item Eccentricity ($e$) 
\item Time of periapse passage ($T_o$), which is when the star comes closest 
to the central dark mass 
\item Inclination (i), which is the angle between the normal to the orbital 
plane and the line of sight and has values ranging from 0 to 180\degr ,
with values less than 90\degr\ corresponding to direct motion 
(position angles increasing with time) and values greater than 90\degr\
corresponding to retrograde motion,
%, and equal to 90$\deg$ corresponding to a straight line
%in the plane of the sky
\item Position angle of the nodal point ($\Omega$), which is the position angle,
measured Eastward of North, of the 
line of intersection between the plane of the sky through the central dark mass
and the orbital plane. In the absence of radial 
velocity measurements (e.g., all stars except S0-2), it is not possible to 
distinguish between the ascending
and descending nodes, which correspond to the nodal points where the
star is moving away from and towards us, respectively, and, by convention, 
the value less than 180\degr\ is taken;  this ambiguity generates a similar 
180\degr\ ambiguity in the longitude of periapse.  
With radial velocity measurements (e.g., S0-2, Ghez et al.\ 2003;
Eisenhauer et al.\ 2004), 
these ambiguities are removed and the ascending node is given for
$\Omega$, with permitted values ranging
from 0 to 360\degr .
\item Longitude of periapse ($\omega$), which is the angle in the plane 
of the orbit, in the direction of motion, from node to periastron, with
permitted values ranging from 0 to 360\degr .

\end{itemize}
In total, this model contains $5 + N \times 6$ parameters, where $N$ is the
number of stars included in the simultaneous fit.  This is a more powerful 
approach than simply averaging the results of N independent orbital analyses,
since each star in a simultaneous solution contributes to the determination
of the common parameters, which in turn leads to a better definition 
of each star's orbital parameters.

\begin{figure}
\epsscale{1.0}
%\plotone{figures/s0srcs.ps}
\plotone{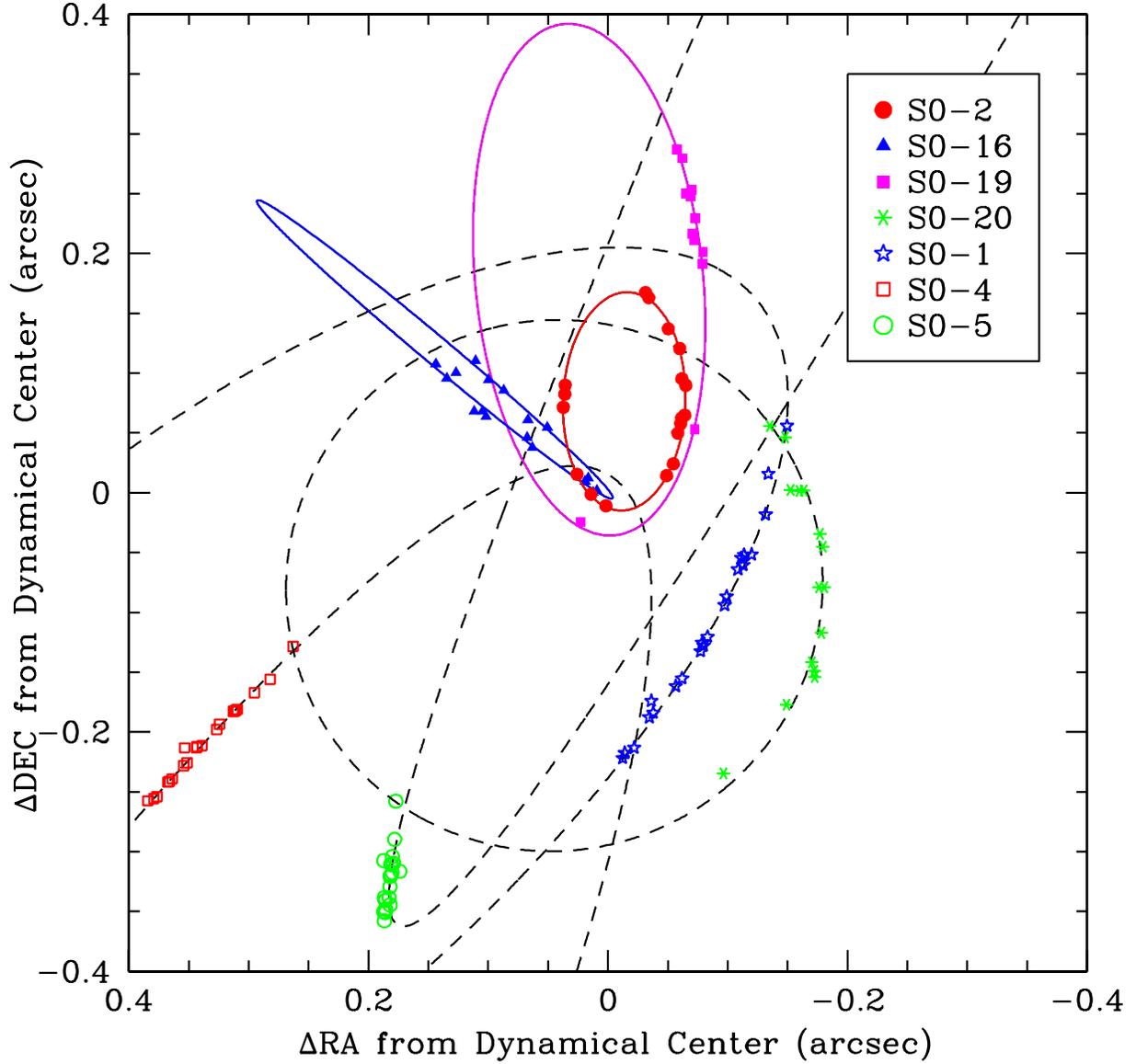}
\figcaption{
Astrometric positions and orbital fits
for the 7 stars 
%within the central 0\farcs 8 $\times$ 0\farcs 8 of the Galaxy,
that show significant deviation from linear motion.
The proper motion measurements were obtained between 1995 and 2003
at the Keck telescopes, have uncertainties that are comparable
to or smaller than the size of the points, and are plotted in the 
reference frame in which the central dark mass' is at rest.
On the plane of the sky, three of these stars show 
orbital motion in the clockwise direction (S0-1, S0-2, and S0-16) and four  
of these stars have counterclockwise motion (S0-4, S0-5, S0-19, and
S0-20).
Overlaid are the best fitting simultaneous orbital solutions, 
which assume that all the stars are orbiting the same
central point mass.  The orbital solutions for the three stars 
that constrain the properties of the central dark object are
delineated by solid lines and the joint orbital solutions for the 
remaining stars are shown with dashed lines. 
\label{gc_pm}}
\end{figure}

The orbital fits, shown in Figure~\ref{gc_pm}, are carried out by minimizing 
the chi-squared value 
between the data and the model and the reported uncertainties are obtained
from the covariance matrix, which corresponds roughly to
changing the total chi-squared values by 1.  
In total, the data set consists of 254 measurements - 126 positional
measurements,  each of 
which provides two independent data points (one for the East-West position 
and the other for the North-South position), and 2 radial 
velocity measurements of S0-2 from a single year, reported by Ghez et al.\ 
(2003).
While the final orbital parameters
reported in Tables~\ref{tab_sol_common} \& \ref{tab_sol_individ} 
come from a simultaneous fit, which is described in detail below,
we first carry out a number of independent 
and semi-independent orbital solutions to 
check the validity of using common values for the central dark mass as well
as to check our estimates of the positional uncertainties.
In these preliminary fits, the central dark mass is not allowed to move
and we scale all the estimated relative position uncertainties by a scale 
factor, which produces a $\chi^2_{dof}$ of 1 for the best independent fits; 
these
scale factors modify the astrometric uncertainties by at most only 30\% and on 
average by only 10\%.
The fully independent solutions yield locations for the central dark mass
that are consistent to within 2$\sigma$,
with individual uncertainties of $\sim$1, 4, and 25 mas 
for S0-2, S0-16 and S0-19, respectively, and larger
uncertainties for the remaining stars.  
Consistency for the central dark
mass is checked by carrying out a semi-independent fit in which the central 
dark object's
location is treated as a common parameter, but its mass is not.
This 
fit is carried out with the 3 stars - S0-2, S0-16, and S0-19 
- that yield meaningful independent mass estimates ($M / \delta M > 3$),
which are consistent to within 2$\sigma$,
with uncertainties of 0.2, 0.6, 1.5 ($\times 10^6$) M$_{\odot}$, 
respectively.
It therefore appears to be well justified to simultaneously fit the
data with a model in which the central dark object's properties 
(M, $r_{RA}$, $r_{DEC}$, v$_{RA}$, and v$_{DEC}$) are 
common to all the stars.  Using an algorithm described
by Salim \& Gould (1999), we solve for the orbital parameters simultaneously
with the inclusion of the central dark object's linear motion on the plane
of the sky as a free parameter.  Since S0-2, S0-16, and S0-19 are the only
stars that have any significant implications for the central dark object's properties,
we divide the problem into two.  A three-star simultaneous fit with S0-2,
S0-16, and S0-19 provides the orbital parameters for these three stars
as well as the central dark objects properties.
The orbital 
parameters for each of the remaining stars are obtained from a
four-star simultaneous fit, which includes the star in question plus 
S0-2, S0-16, and S0-19; this was done to appropriately include the effects 
of the uncertainties in the central dark object's parameters in estimates
of the remaining stars' orbital parameters.  The resulting 
$\chi^2_{dof}$ for all the
simultaneous fits are comparable to 1, again supporting the use of a point 
mass potential model.

\subsubsection{Orbital Fit Results}

Estimates of the central dark mass' properties from the three-star simultaneous
fit are reported in Table~\ref{tab_sol_common}.  The central dark mass is estimated to be
$3.7 (\pm 0.2 ) \times 10^6 (\frac{R_o}{8kpc})^3 M_{\odot}$.
While this is consistent with that inferred from the 
orbit of S0-2 alone (Ghez et al.\ 2003; Sch\"odel et al.\ 2003), its 
uncertainty is a factor of 3-4 times smaller due, primarily, to
the longer time baseline for the measurements, and, 
in part, to the additional information
offered by S0-16 and S0-19.
This makes distance, which is fixed in all the orbital analyses reported 
thus far, the limiting uncertainty for the first time (see also Eisenhauer et 
al. 2004); 
the 0.5 kpc uncertainty in the Galactic center distance (Reid 1993)
contributes an additional 19\% uncertainty in the estimated mass, beyond that
reported in Table~\ref{tab_sol_common}.
Similarly to the mass, the inferred center of attraction agrees well with 
the results from the analysis of S0-2's orbit by Ghez et al.\ (2003).
The location is only modestly
improved in the simultaneous fit, because 
the black hole's proper motion is treated as an unknown variable only in 
the multiple star orbit model, which increases the formal uncertainties 
in the black hole's location.
The estimate of the
dark mass's motion on the plane of the sky is the first such estimate
derived from orbital fits.  While a single star's
orbital trajectory can, in principle, constrain this motion, 
in this solution it is primarily constrained by the closest approaches 
of S0-2, S0-16, and S0-19 and their span of periapse passage times of 5 years.  The inferred 
proper motion of the dark mass, with respect to the central 
stellar cluster, is 1.4$\pm$ 0.5 mas yr$^{-1}$, statistically 
consistent with no motion.  Overall, simultaneously fitting the stellar orbital 
motion has allowed significant improvements in the derivation of the
central dark object's properties.
 
With the central parameters constrained simultaneously by multiple stars,
the precision with which each star's orbital elements can be determined
is also greatly improved compared to that obtained from an independent orbit 
analysis.  Table~\ref{tab_sol_individ} lists the parameters specific to the 
individual stars from the simultaneous fit.
Over the course of this study (1995 - 2003), these stars have
either undergone periapse passage or are remarkably close to periapse.
The smallest periapse distance is achieved by S0-16, which comes
within 45 AU with a velocity of 12,000 $\pm$ 2,000 km s$^{-1}$.
%Similarly, the
%orbits' orientations are remarkably well determined for stars that in some
%cases have undergone only a small fraction of their orbit.  In these cases,
%this is possible because other stars are defining the central mass' properties.

\begin{figure}
\epsscale{1.0}
%\plotone{figures/Ae.ps}
\plotone{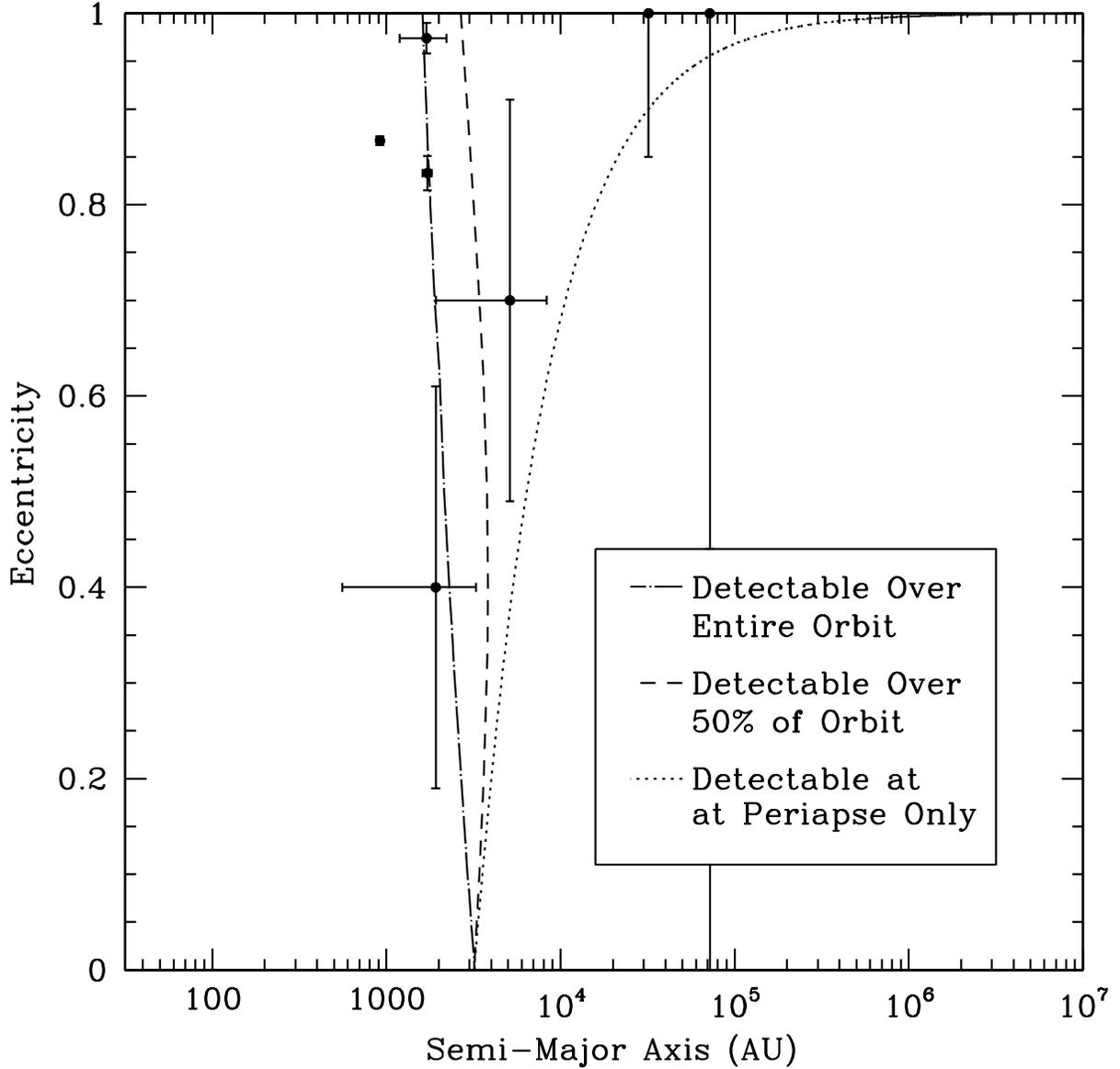}
\figcaption{Eccentricity vs. Semi-Major Axis  
for the 7 stars included in this study.
The observational selection
effects are quantified by the fraction of time a face-on orbit experiences 
acceleration larger than our threshold value of 2 mas y$^{-2}$, with the cases 
of 100\% of the orbit (long dashed - dotted line), 50\% of the orbit (short dashed 
line), and only periapse passage (dotted line) shown.  
The region to the left of the long dashed - dotted line is free of
observational selection effects and therefore should not be missing
any stars that are brighter than $\sim$15.5. 
\label{gc_ae}}
\end{figure}

There are clear selection effects in this study that must be understood and
accounted for before the ensemble properties of the sample can be studied.
Since a star has to experience acceleration in the plane of the sky
of greater than 2 mas y$^{-2}$
to be included in the orbital analysis,
there is an observational bias towards detecting stars in eccentric
orbits at periapse,
in spite of the fact that a star spends most of its time 
away from periapse.
Stars experience their largest acceleration near periapse, at a projected 
distance which scales as $q = A(1-e)$.  For a given semi-major axis 
above $\sim$3200 AU, this allows stars in highly-eccentric orbits
to have detectable accelerations near their closest approach, while 
stars on low-eccentricity orbits will be 
below the detection threshold in all parts of their orbits.
Figure~\ref{gc_ae} quantifies these effects based on the fraction of time a
face-on orbit experiences accelerations larger than our threshold value. 
Four stars -- S0-2, S0-16, S0-19, and S0-20 -- lie in the parameter space 
that is unbiased, 1 star --  S0-1 -- resides in a region that is mildly biased
($\sim$50\% effect), and the last 2 stars -- S0-4 and S0-5 -- are detected
only because they are on eccentric orbits and remarkably close to periapse 
passage.
The excess of high eccentricity orbits in our sample is therefore a consequence 
of an observational 
bias; restricting the analysis to the 5 stars for which the bias is 
$\lesssim$50\% effect, we find 2 stars with eccentricities $\lesssim$0.70,
which is statistically consistent with isotropy, if we assume that an
isotropic system has a cumulative probability distribution $\propto e^2$
(Binney \& Tremaine 1987).   

The distribution
of semi-major axes is also noteworthy.  While there are no
observational selection effects against it, there is a distinct lack of
stars as bright as those tracked in this study (K $\lesssim$ 15.5) having
semi-major axes $\lesssim$1000 AU, and likewise apoapse distances of 
$\lesssim$1800 AU.
The other end of these distributions, however, are not seen due to selection
effects.  

\begin{figure}
\epsscale{1.0}
%\plotone{figures/incl_map.ps}
\plotone{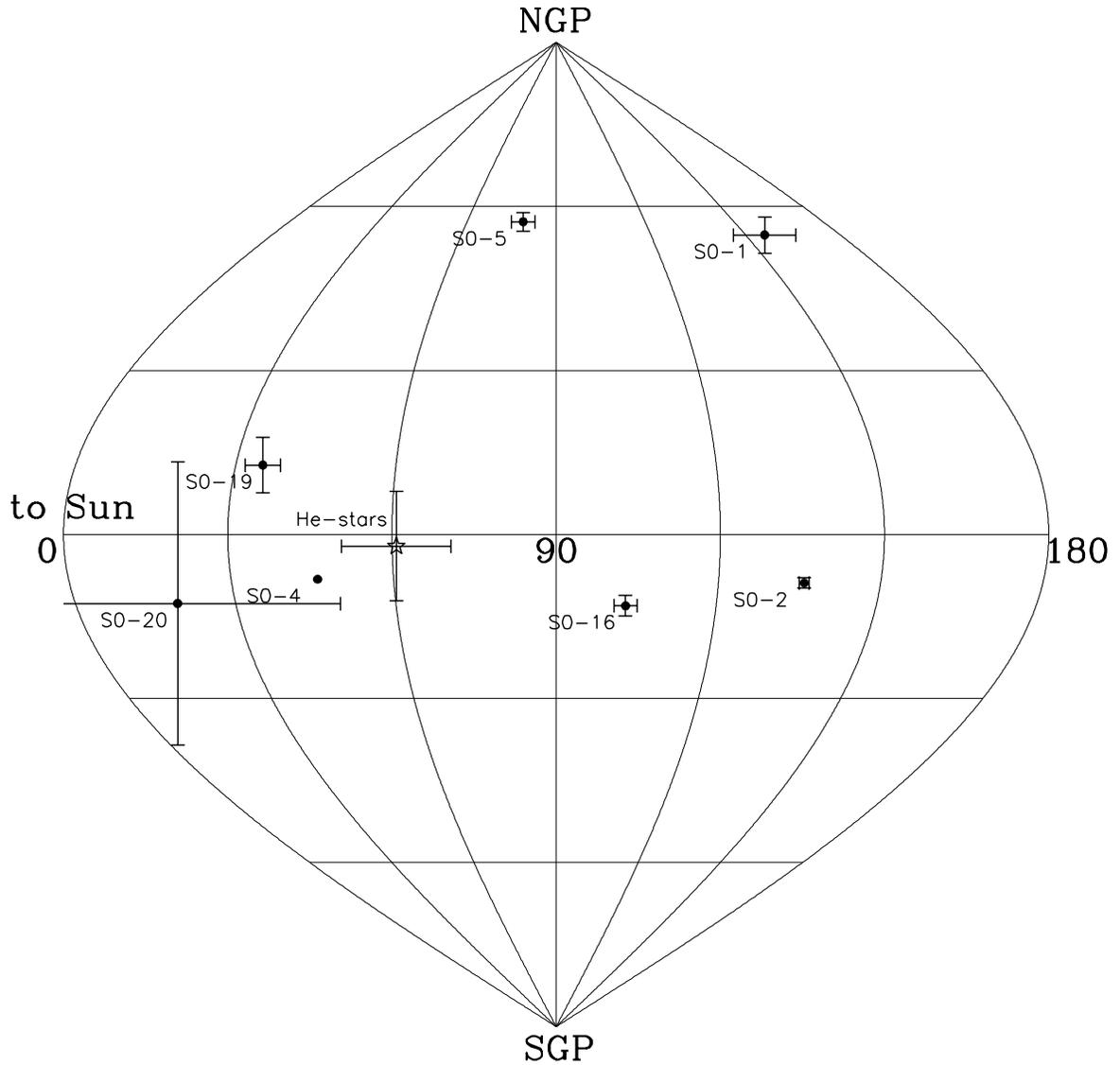}
\figcaption{
Positions of the angular momentum vectors for an observer at the center of the
Galaxy.  Only one hemisphere is shown (East as seen from the Earth), 
since the degeneracy of the inclination sign makes it impossible to know which 
hemisphere a vector points, except for S0-2.  The large uncertainties for S0-4
have been omitted for clarity.  If the orbits were to be co-planar
the angular momentum vectors would cluster, which we do not see.  The normal to
the plane of the He~I stars found by Levin \& Beloborodov (2003) is also 
indicated.
\label{gc_angmom}}
\end{figure}

\begin{figure}
\epsscale{1.0}
%\plotone{figures/apoc_map.ps}
\plotone{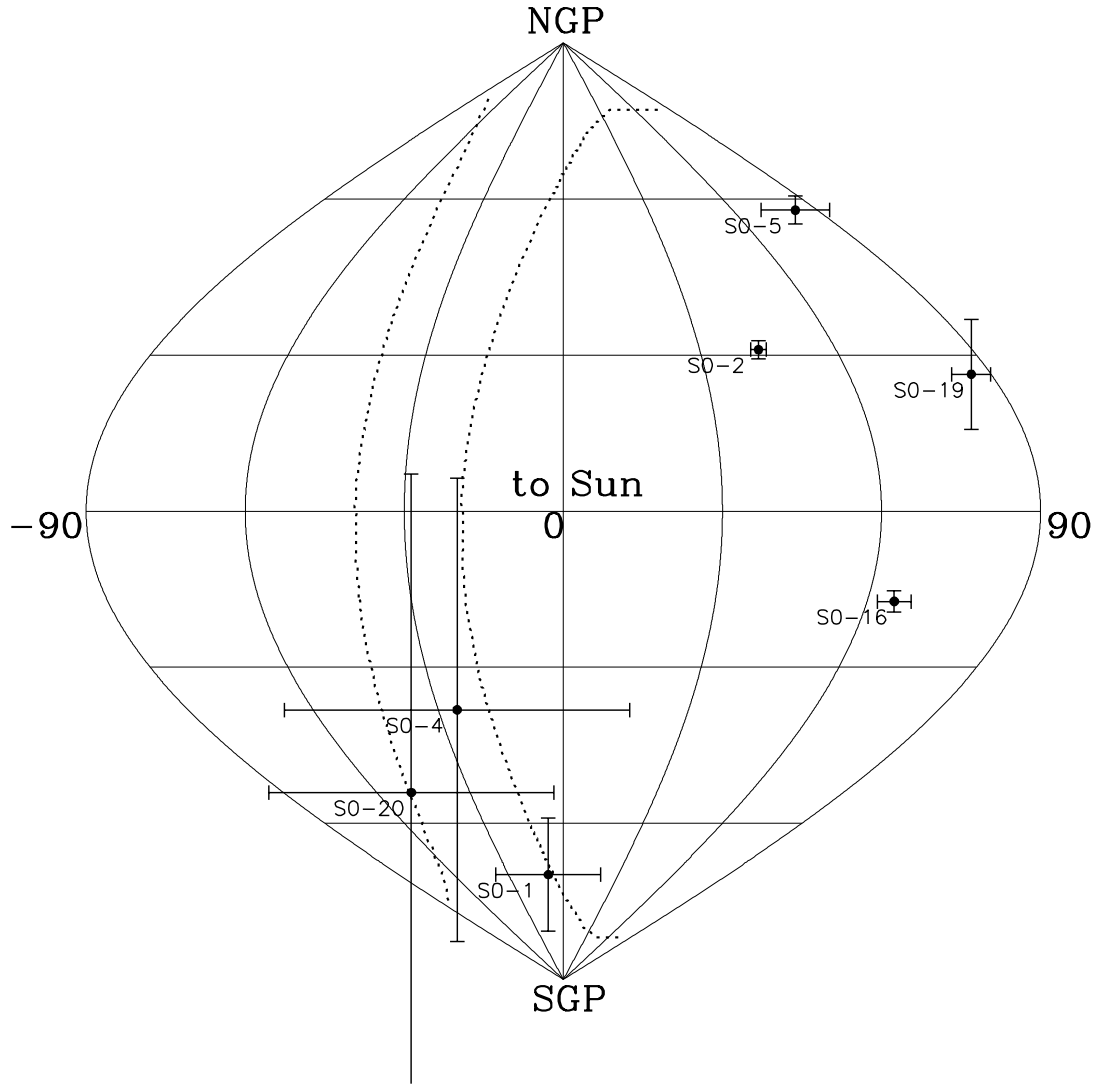}
\figcaption{
The directions of the apoapse vectors, 
as seen from an observer at the Galactic center, 
for the hemisphere containing S0-2.  
%Degeneracy in the inclination angle
%for the other stars can not re-position the vectors in this hemisphere. 
The He~I star plane of Levin \& Beloborodov (2003) is shown as a 20\degr\
wide band (dotted lines).  These vectors are consistent with an isotropic distribution,
even when the degeneracy of the inclination sign is taken into 
account.
\label{gc_apoapse}}
\end{figure}

In contrast to the shape of the orbit, its orientation
should be unaffected by observational bias.
To fully describe the orientations of the orbits, it is necessary
to specify the directions of two vectors, one normal to the orbital plane, such 
as the angular momentum vector, and one along the semi-major axis, such as
the direction to apoapse.  
For S0-2, which has both astrometric and radial velocity measurements, 
the full three-dimensional orbit is unambiguously determined.  For the
stars with only astrometric measurements available, however, a 
degeneracy exists from a possible reflection about the plane of the sky;
we therefore assume
that these orbits are oriented such that the unit angular momentum
and apoapse direction vectors 
are in the same hemisphere as those quantities derived for S0-2.
Figures~\ref{gc_angmom} and \ref{gc_apoapse} show the directions for these 
unit vectors.  
The directions of neither the angular momentum nor apoapse 
vectors show any clear preferred direction or co-planarity,
and are statistically
consistent with a random distribution of orbits.

\section{Discussion}

\subsection{The Case for and Properties of the Central Supermassive 
Black Hole}

Stellar orbits provide the most direct measure of the amount of dark
matter concentrated at the center of the Galaxy.  
Compared to masses inferred from the velocity dispersion measurements,
the mass derived from multiple orbits, $3.7 (\pm 0.2) \times 10^6 M_{\odot}$,
is a factor of 2 higher than that
estimated by a non-parametric approach presented by Chakrabarty \&
Saha (2001), which is supposed to be the most robust approach, and is somewhat less discrepant with the parametric approaches
(e.g., Ghez et al.\ 1998; Genzel et al.\ 2000; Sch\"odel et al.\ 2003).  
Since the mass estimates from the velocity dispersion measurements and orbital 
fits have all assumed the same distance and all depend on distance 
as ${R_o}^3$, the assumptions about distance are not the source 
of this mass discrepancy.  By simultaneously solving for multiple orbits, 
we now have only one more parameter left out of the fit for a Keplerian orbit 
model, that of the black hole's motion along the
line of sight.  Given the small values for its motion on the plane of the sky,
this last parameter is unlikely to have any significant effect on the estimated
mass.  The two possibilities therefore lie in problems with the mass estimates
from the velocity dispersions.  First, the mass estimates could be biased 
by the weighting schemes used to calculate the velocity dispersions.
While roughly 100 stars have reported proper
motion values in the earlier works of Ghez et al.\ 1998 and Genzel et al.\ 2000,
only 18 of these have S/N $>$ 5 and half of them have S/N $<$ 3, making the 
velocity dispersion bias term non-negligible.
Second, the projected mass estimators could be biased by the properties 
of the central stellar cluster.  Specifically, both the level of anisotropy
and the slope of the stellar density distribution can significantly alter
the values inferred from standard projected mass estimators (Genzel et al.\
2000; Figer et al. 2003).  Different results have been reported both for the presence of
anisotropy (Ghez et al.\ 1998; Genzel et al.\ 2000; Sch\"odel et al.\ 2003)
and the radial distribution of stars (Scoville et al.\ 2003; Genzel et al.\ 
2003a; Figer et al.).  While a full exploration of these effects is outside the scope of 
this paper, here we emphasize that the orbital mass is more robust
and should be used in all future characterizations of 
the Galaxy's central dark mass concentration.

Stellar orbits confine the central dark mass of 
$3.7 (\pm 0.2 ) \times 10^6 (\frac{R_o}{8kpc})^3 M_{\odot}$  to
within 45 AU, the closest approach of S0-16, implying a minimum 
density of $8 \times 10^{16} M_{\odot} / pc^3$ for the central dark mass.
This confines the mass to a volume that is a factor of 20 smaller 
than that inferred from S0-2 (Sch\"odel et al.\ 2002; Ghez et al.\ 2003) and
increases the inferred density by 
four orders of magnitude compared to 
estimates from measurements of acceleration vectors (Ghez et al.\ 2000;
Eckart et al.\ 2002).
At this density, the two existing alternative explanations to a 
supermassive black hole
for the compact dark object found at the center of the Galaxy
become significantly less tenable (see also Sch\"odel et al.\ 2002).  Any cluster of dark objects, such as 
those considered by Maoz (1998), would have a lifetime of a 
mere $\sim$10$^5$ years, owing to gravitational instability, which is 
significantly shorter than the age of the 
Galaxy, making this a highly unlikely explanation for the central dark 
mass concentration.  
For the fermion ball hypothesis (Viollier et al.\ 1993), the mass of the
constituent particles is now required to be 
$74 keV c^{-2} (\frac{0.3}{R})^{3/8} (\frac{2}{g})^{1/4} (\frac{3.7 \times 10^6}{M})^{1/8}$,  where R and M are the radius in milli-pc and mass in M$_{\odot}$ 
of the fermion ball, 
respectively, and g is the spin degeneracy factor of the fermion; 
this is 5 orders of magnitude larger than the current limits on degenerate
neutrino species (Spergel et al.\ 2003), rendering the fermion ball hypothesis
also highly unlikely (see also discussion in Sch\"odel et al.\ 2002).   
With the Galaxy's central dark mass now confined to a 
radius equivalent 
to 600 $\times$ the Schwarzchild radius of a $3.7 \times 10^6 M_{\odot}$ 
black hole, the multiple stellar orbits present the strongest case yet for a 
supermassive black hole at the center of the Milky Way galaxy.

The measured linear velocity of the central black hole on the plane of the sky 
limits the mass of any possible companion black hole, through the assumption
that any velocity is due to reflex motion.
With a 1 $\sigma$ upper limit of 2 mas y$^{-1}$, the mass of any possible 
companion black hole is constrained to be less than 
$\sim 5 \times 10^5 (R / 16,000 AU)^{1/2} M_{\odot}$, where $R$ is the
distance of the companion black hole from the central black hole; the 
generalization of this limit to other radii works as long as 
the black hole companion lies outside the orbits that 
contribute to the determination 
of the central dark object's properties (Table~\ref{tab_sol_individ}) 
and that the orbital period is long compared to the duration of the 
study, 8 years.  A related measurement comes from upper limits inferred
for the motion of the radio source Sgr A*, which is
assumed to be associated with the central black hole (see discussion in
\S4.2).  In the plane of the Galaxy, the upper limit is 
$\sim$ 20-25 km s$^{-1}$ (Backer \& Sramek 1999; Reid et al.\ 1999;
Reid \& Brunthaler 2004, which is comparable to our limits.
Perpendicular to the Galaxy, Reid \& Brunthaler (2004) derive
a more constraining upper limit 
of $\sim$1 km s$^{-1}$.
Nonetheless, the infrared and
radio measurements are fundamentally different.  In the infrared, the
black hole's motion is measured with respect to the central stellar cluster,
which traces the local barycenter, while in the radio, Sgr A*'s motion is
derived with respect to background quasars, so motions of the black hole 
along with the central stellar cluster as well as the solar parallax 
show up in this measurement. 
Therefore, while the radio upper 
limit on the motion of Sgr A* is smaller than the infrared upper limit 
on the motion of the black hole, the latter is a more direct measure of the
upper limit for the reflex motion from a possible black hole companion.
%imposes a mass limit that is a factor 
%of 4 times smaller than that derived from the orbital motions. 
%Therefore an equal mass companion black hole outside
%the orbits of these stars is ruled out, however a lower mass companion
%is permissible on the basis of the central mass's reflex motion.

In the context of other galaxies, the Milky Way's central dark mass 
concentration
distinguishes itself in terms of both its inferred density and
mass.  The Galaxy's central minimum dark mass density now 
exceeds the minimum dark matter density inferred for 
NGC 4258 (Greenhill et al.\ 1995; Miyoshi et al.\ 1995) by five orders of 
magnitude, reinforcing the Milky Way as the strongest case for a black hole
at the center of any normal type galaxy.

It is also possible to use the observed dark mass concentration in
the Milky Way to further explore the fermion ball
hypothesis as a universal alternative explanation for 
supermassive compact objects in all galaxies as has been 
proposed in the past.
For objects
composed of the minimum mass particles imposed by the stellar orbits
in the Galactic center, the maximum mass is
$1 \times 10^8 M_{\odot} (\frac{76 keV}{mc^2})^2 (\frac{2}{g})^{1/2}$,
from the Oppenheimer-Volkoff limit 
(Munyaneza \& Viollier 2002).  This 
is less massive than half of the 
supermassive compact objects that have been identified thus far
(cf., for example, the compilation in Tremaine et al.\ 2002), thereby eliminating
an all-encompassing fermion ball hypothesis.

In contrast to its high minimum central dark mass density, the Milky Way 
appears to harbor the least known massive supermassive black hole, as inferred
directly from dynamical measurements.  It therefore 
potentially has an important role to play in assessing the
M$_{bh}$ vs. $\sigma$ relations (e.g., Ferrarese \& Merritt 2000; Gebhardt et al.\ 2000).  
However, the current impact of the Milky Way on the
M$_{bh}$ vs. $\sigma$ relation is limited by uncertainties in the
determination of its bulge velocity dispersion (Tremaine et al.\ 2002).  Nonetheless, the higher mass value 
from the orbits brings our Galaxy
into better agreement with the M$_{bh}$ vs. $\sigma$ relationship derived
from a large sample of galaxies (e.g., Tremaine et al.\ 2002; Merritt \& 
Ferrarese 2001).

\subsection{Sgr A* and other Possible Counterparts to the Central Black Hole}

The orbits provide very precise information on the location and 
kinematics of the central supermassive black hole, allowing us to 
explore its association with the radio source Sgr A* and any
possible near-infrared counterparts.  
In Appendix B, we derive the infrared position of the radio source 
Sgr A*.
Relative to the dynamically determined position of the central dark mass, 
which is known to within 1.3 mas (1 $\sigma$),
the inferred infrared position of Sgr A*, which is less accurately known -
is offset by a mere 0.5 $\pm$ 6.4 mas W; 
and 9 $\pm$ 14 mas S; 
the two positions therefore appear to be consistent
to within 1$\sigma$.
%The upper limit on the black hole's motion, $<$30 km s$^{-1}$,
%also agrees well 
%with the upper limit inferred for the motion of Sgr A*
%$<$ 8 km s$^{-1}$ (Backer \& Sramek 1999; Reid et al.\ 1999; Reid et al. 
%2003b) 
%\footnote{It is however worth noting that
%the upper limit on the black hole's motion inferred from the stellar orbits, 
%which is determined with respect to the stellar cluster,
%is a more direct measurement than that derived from the radio measurements
%of Sgr A* with respect to background quasars,
%which require a model of the solar reflex motion to be removed.}.  
Furthermore, using the kinematics of S0-2 from Ghez et al.\ (2003)
and the upper limit on the motion of Sgr A*, Reid et al.\ (2004) argue 
that Sgr A* has a minimum mass of $4 \times 10^5 M_{\odot}$, consistent
with the black hole mass estimated from orbital motion. 
Given the agreement in position, velocity (discussed in \S4.1), and mass, 
it appears that Sgr A* is indeed associated with the black hole at the 
Galaxy's center.

Identifying near-infrared counterparts to the central black hole is
a difficult task, given the high stellar
densities, velocities, and accelerations
at that location.  S0-19 serves as a good 
illustration of these challenges. 
Its large proper motion and strong curvature in a crowded region 
makes it challenging to track and led Genzel et al.\ (1997) to propose 
their 1996.43 detection of this source (their label S12) as the best candidate 
for the infrared emission from the central black hole; at that time,
this source was coincident to within
1$\sigma$ (30 mas) of the relatively crude position of Sgr A* reported by Menten et al.\ (1997).
With the newly determined location of the black hole based on orbits, 
it is now clear that this source is offset by 54 mas, or 41$\sigma$,
from Sgr A* and that it is simply one data point in the trajectory of
the high velocity star S0-19 that, in 1996, was near the black hole\footnote{S0-19 was detected by Ghez et al.\ (1998) in 1995 with two 
possible counterparts identified in 1996.  With limited time coverage, Ghez et 
al.\ (1998) were not able to definitively identify either as the correct 
counterpart to either S0-19's 1995 position or Sgr A* and therefore 
did not include this source in their proper motion sample.}.

The search for infrared counterparts to the central black hole is greatly
facilitated by the use of stellar orbital motions to refine its location
by a factor of 20 compared to Menten et al.\ (1997) and a factor of 8 compared
to Reid et al.\ (2003).  During 4 of the 9 years of this study, a star
with measurable proper motions is detected 
within 54 mas of this location 
%In 1995 and 2000 - 2002, a bright star was located 
%within 50 mas of this location, 
preventing a faint counterpart from being 
easily detected (S0-19 in 1995, S0-16 in 2000, and
S0-2 in 2001-2002). 
S0-21 (K=16.1 mag) 
is the only source in this study without unambiguous proper motion
and its 3 measurements are all within 3$\sigma$ of the
black hole's location; with only a 1 year time baseline (1998.25-1999.56) 
its proper motion is less than 22 mas y$^{-1}$ (1$\sigma$).
While this could be a counterpart, we believe that it is not. 
There are 2003 correlation peaks that do not pass our 3-submap requirement,
but, if real, indicate that S0-21 has measurable proper motion 
over this longer time baseline.
In the remaining 3 years, 1996-1997 and 2003,
there is no source detected by the relatively conservative source 
identification criteria set forth in \S3.1  
within 3 $\sigma$ of the dynamically determined location of the black hole.
We therefore infer that no steady source brighter than 
$\sim$15.5, the magnitude of the faintest star we were able to 
identify in this region in all epochs without any a priori-information 
(see \S3.1), was coincident with our
inferred black hole position during our observations 
(see also Hornstein et al.\ 2002, 2003)\footnote{We note that after submission of this paper, 
a variable source coincident with Sgr A* was detected at near-infrared 
wavelengths (Genzel et al.\ 2003b; Ghez et al.\ 2004); its characteristics are 
consistent with our non-detection of a steady source brighter than 15.5 mag}.

\subsection{The Origin of the Central Stellar Cluster}

The orbital parameters derived here provide important clues for 
understanding the origin of the Sgr A* cluster stars, which appear to 
have hot photospheres similar to those of massive young stars (Genzel
et al.\ 1997; Eckart et al.\ 1999; Figer et al.\ 2000; Gezari et al.\ 2002;
Ghez et al.\ 2003).
In the context of the luminous He~I emission line stars,
which are located an order of magnitude further from the black hole than the
Sgr A* cluster stars,
several ideas have been proposed to account for 
apparently young stars in a region whose current conditions seem to be 
inhospitable to star formation:
1) that these are old stars 
masquerading as youths, 2) that they 
were formed more or less {\it in-situ} by a cataclysmic compression of an 
already dense cloud or disk, and 
3) that they were formed elsewhere as part of a massive cluster, but migrated 
inwards rapidly by dynamical friction.  Here, we 
briefly examine each of these hypotheses in the context of
the Sgr A* cluster stars.

\subsubsection{Old Stars Masquerading as Youths}

Stellar mergers of relatively old stars can, if the stellar density is
sufficiently large,
produce stars massive enough to appear as main-sequence OB stars.
This scenario is likely to produce stars whose orbits are isotropically
distributed, consistent with our observations.
However, there are several challenges to this hypothesis for the stars
in the Sgr A* cluster.  First, several successive mergers of stars of
increasing mass are required to produce a star resembling S0-2 ($\sim$O9.5,
$M \sim 15 M_{\odot}$; Ghez et al.\ 2003), unless the mass segregation in this region has been so
strong that only stars $\gg 1 M_{\odot}$ are left.
Second, as the merger products become more massive, their nuclear lifetimes
decrease, so that there is less time available for the next merger event
in the sequence.
Using a Fokker-Planck approach, Lee (1996)
investigated the stellar merger hypothesis for the massive emission-line stars
in the central parsec, and concluded that an insufficient number of them is
likely to be present.
The Sgr A* cluster stars, however, are much more
concentrated toward the center where the stellar density is maximized
(Genzel et al.\ 2003a)
and the collision time is correspondingly
shorter, so in this respect, merger events may be relatively favored there.
However, the third challenge is that the velocity dispersion of stars
near the supermassive black hole, 400 km s$^{-1}$ at 0.01 pc 
(e.g., Ghez et al.\ 1998), 
is comparable to the escape velocity from the surface of a main-sequence
O9.5 star, $\sim$1000 km s$^{-1}$, so collisions in the volume occupied
by the Sgr A* cluster stars are therefore less likely to lead to mergers
and mergers that do occur are likely to be accompanied by significant mass
loss (Freitag \& Benz 2002).  
A fourth consideration which may disfavor the collisional mechanism
is the relatively normal rotation rate of S0-2 (Ghez et al.\ 2003).
Alexander \& Kumar (2001) have found that tidal encounters between
main-sequence stars in the central cluster can eventually spin up
those stars to near break-up speed.  Colliding stars effectively
represent an extreme example of this phenomenon, so merger products
should be much more rapidly rotating than S0-2 appears to be.  Of
course,
the apparent rotation rate of S0-2 can be attributed to a particular,
low-probability orientation, so the measurement of absorption-line
widths in just one additional member of the cluster should clarify
this point.  
%A further stochastic indicator of merger history is the
%distribution of orbital eccentricities of the SgrA* cluster stars.  A
%succession of mergers is likely to favor low-eccentricity orbits, so it
%is not yet clear whether the biased distribution of relatively eccentric
%orbits can be consistent with the merger scenario.  
While further calculations
are clearly required to assess the importance of this complex mechanism,
at present it appears to be quite unlikely.

Another suggestion to account for the Sgr A* cluster stars without invoking
star formation is that they may be exotic objects.  This catch-all category
includes a number of possibilities.  For example, it is reasonable to expect
that stellar remnants such as neutron stars and black holes sink into the
central few milli-parsec as a result of dynamical mass segregation
(Morris 1993; Lee 1996; Miralda-Escud\'e \& Gould 2000).  Mergers of these
remnants with normal stars could produce Thorne-Z\'ytkow objects, or,
in the case of black hole remnants, something with dubious long-term
stability.  However, Thorne-Z\'ytkow objects are expected to appear as
red giants or supergiants rather than massive blue stars, and
may be unstable;
%(brad suggested some references that I can't read here) 
likewise,
stable stellar objects with black hole cores have yet to be described.
If the stellar remnant that undergoes a merger is a white dwarf, then a 
``reborn star" results, and it could be suitably luminous.
However, such an object would probably be a red giant rather than an
early-type star, and in any case, the white dwarf precursor is likely
to have migrated out of the central region because of its low mass.
Another, slightly less exotic possibility for the Sgr A* cluster stars
is that they be the exposed, hot cores of tidally stripped red giant
stars.  Indeed, there appears to be a paucity of red giants in the inner
0.2 pc of the Galaxy (Sellgren et al.\ 1990; Genzel et al.\ 1996), 
suggesting that red giant atmospheres are collisionally removed there,
possibly by collisions with binaries (Davies et al.\ 1998).  However, the 
luminosity of the exposed stellar cores may be too small to account for
the Sgr A* cluster stars (e.g., Sch\"onberner 1981, 1983).
%[XXX should we allow for smaller "stripping" events?]

\subsubsection{Recent In-Situ Star Formation}

The second category of hypotheses is that the early-type stars really did form 
recently {\it in situ}.  To do this, the parent cloud would have to have 
undergone violent compression to densities exceeding the limiting Roche 
density.\footnote{Tidal compression may be a contributor: a cloud moving 
toward the center on a purely radial trajectory will experience a compression 
in two-dimensions, although this would be partially counteracted by distension in the 
radial dimension, so that the net compression would not be a strong function 
of radius, and is not likely, by itself, to be able to raise the density by 
the many orders of magnitude necessary.  In addition, any non-radial motion 
would imply a tidal shear in the azimuthal direction which would
also counteract the tidal compression.}  This hypothesis warrants 
consideration because the mechanism for violent compression of any cloud 
passing close to the black hole is innate to the model.  A dense cloud 
brought within 0.02 parsecs of the supermassive black hole would unavoidably 
lead to a high rate of accretion onto the black hole.  
If the onset of this accretion is rapid, the resultant release of accretion 
energy would be powerful enough to compress the cloud.  
%[somewhere I wrote down a 
%relevant reference on how much a cloud can be compressed].  
Morris, Ghez \& Becklin (1999) proposed that this process can manifest itself 
as part of a limit cycle involving the circumnuclear disk (CND).  This disk 
currently has a central cavity of 1 parsec radius, presumably because of the 
outgoing ram pressure of the winds from the cluster of luminous, early-type stars in the 
central parsec.  However, as the lifetimes of these stars is $\sim$10$^7$ 
years, and because the CND 
itself undergoes viscous evolution on times more comparable to the orbital 
time at the inner radius, $\sim5~\times~10^4$ years, the inner edge of the 
CND will migrate toward the central black hole on a time scale comparable to 
the stellar 
evolution time.  When the first portions of the CND reach the central black 
hole, the outgoing shock resulting from the accretion event provokes massive 
star formation in the now nearby disk by strong compression.  The strong winds 
from these stars cause the inner disk boundary to recede 
and the cycle begins anew.  
A weakness of this hypothesis is the magnitude of the required compression.
While the density of a cloud which has migrated close to the black hole might be
substantially larger than the densities so far inferred for any of the gas in 
the region, it is difficult to see how even the most effective compressive 
event can bring gas up to the limiting Roche density.  

\subsubsection{Recent Star Formation at Large Galactic Center Distance 
Accompanied by Rapid Orbital Migration}

The third hypothesis which has been considered is that the early-type stars 
in the central parsec formed well outside the central parsec, but migrated
inward under the action of dynamical friction on time scales substantially less
than their nuclear time scale, $\sim$10$^7$ years.  This is not possible for
individual stars (Morris 1993), but Gerhard (2001) has pointed out that, because
the dynamical friction time scale is inversely proportional to an object's mass,
sufficiently massive clusters can migrate to the central parsec from radii of 
tens of parsecs within the required time, especially if they remain bound to
their parent cloud.  This hypothesis has been investigated numerically by 
Kim \& Morris (2003) and 
Portegies-Zwart, McMillian \& Gerhard (2003), who clarify that very massive 
clusters are 
required --
10$^{5-6}$ M$_{\odot}$, far more massive than even the extreme (for our
Galaxy) Arches and Quintuplet clusters (e.g., Figer, McLean, \& Morris 1999; 
Figer et al.\ 2002).
Core collapse is inevitable in the massive, dense clusters required
for the cluster inspiral hypothesis.  This process helps ensure that,
in spite of tidal stripping of stars outside the cluster core as the
cluster migrates inward, there remains a tightly bound, cluster core 
that survives intact into the central parsec.  However, Kim \& Morris 
find that the 
mass of stars reaching the central parsec, for any feasible initial cluster 
mass, substantially exceeds the mass of early-type stars in the central parsec 
cluster.  

More recently, Portegies-Zwart
and McMillan (2002, see also Rasio, Freitag, \& G\"urkan 2004) have 
raised the possibility that core collapse in sufficiently massive clusters proceeds all the way to the formation of an intermediate-mass black hole (IBH), which can carry cluster stars in with it as it
spirals inward by dynamical friction.  The implications of such a
cluster-produced IBH for the distribution of early-type stars in the central parsec have recently been investigated by Hansen \& Milosavljevi\'c (2003).  They argue that the HeI emission-line stars in the central 
parsec have been tidally stripped from the IBH during successive passages
near the supermassive black hole, but that they retain a memory of 
the IBH orbit.  
Furthermore, cluster evaporation during
the inspiralling process leads to a marked decrease in the effectiveness of 
this process; a remnant core of an initially globular-cluster-mass cluster can 
reach the central parsec only by distributing a large number of early-type stars
at all radii, whereas there is currently no evidence for a young population 
beyond the central parsec.  
In addition, Kim, Figer, \& Morris (2004) find that an IBH helps deliver
stars to the central parsec only if it contains at least 10\% of the 
cluster mass, far larger than masses obtained in simulations of 
successive merger (Portegies-Zwart \& McMillan).
Further investigations of this hypothesis are warranted, though
it currently appears to suffer from a number of difficulties.

Both the {\it in situ} formation mechanism and the evaporating,
inspiralling cluster mechanism will lead primarily to a disk of stars,
the first because the inwardly migrating reservoir of gas inevitably
forms a disk by virtue of its angular momentum, and the second because
the stars lost from the cluster will retain a memory of the direction
of the cluster angular momentum.  While most of the early-type
emission-line stars in the central parsec appear to orbit in or near a
well-defined plane (Levin \& Beloborodov 2003), the stars in the SgrA*
cluster do not.  Levin \& Beloborodov argue that the SgrA* cluster
stars and the more distant emission-line stars all formed at about
the same time in a starburst taking place in a thick accretion disk
around SgrA* (see also Nayakshin \& Cuadra 2004).  Unlike the He-I emission line stars, the orbits of the
SgrA* cluster stars are likely to have 
been altered by Lens-Thirring precession caused by the massive central
black hole, so that their orbital angular momenta vectors should
form a plane, which is inconsistent with the observations (see Figure 3).  
This suggests that the Sgr A* cluster stars were not formed by these
mechanisms, which produce an initial common direction for the 
angular momentum vector.  

One alternative hypothesis for the tight orbits of the SgrA* cluster
stars is that they have resulted from the tidal disruption of massive star
binaries as stars presumably related to the HeI emission-line stars
undergo relatively close passages by the supermassive black hole 
(Gould \& Quillen 2003).  These authors estimate that a sufficient 
number of single stars resembling S0-2 can be scattered onto orbits similar
to those of the SgrA* cluster stars to explain that cluster, if they
originate in binary systems undergoing close passage by the black hole.  Multiple encounters with other stars in this region are required to 
bring the apoapse distances down to the range of values exhibited by the
SgrA* cluster stars, i.e., far smaller than the typical orbital radii of 
the more massive emission-line stars.  It remains to be seen whether
this hypothesis can account for the SgrA* cluster stars. 

In sum, there are serious difficulties or open questions associated with all 
of these hypotheses, although few of them can be definitively ruled out.  
While the stars with known orbits offer modest support for hypotheses
that produce isotropic distributions, this is based on a very small
sample.  Additional orbits for stars in the vicinity of the central black hole
may ultimately provide a 
sufficiently strong constraint to cull this list of possibilities.  
In the meantime, we are left with an interesting conundrum.

\section{Conclusions}

After almost a decade of diffraction-limited imaging at the W. M. Keck I 
10-meter telescope,
we have obtained orbital solutions for multiple stars.  
This orbital analysis has the
advantage of simultaneously solving for a common set of properties for the
central dark object, which not only reduces the uncertainties 
in the black hole's mass and location compared to an analysis that treats
each star independently, but also provides the first direct measure of the 
black hole's velocity with respect to the central stellar cluster.
Together, the stellar motions reveal a 
central dark mass of
$3.7 ( \pm 0.2 ) \times 10^6 (\frac{R_o}{8kpc})^3 M_{\odot}$ and
confine it to within a radius of
a mere 45 AU or equivalently 600 R$_{sh}$, dramatically strengthening
the case for a supermassive black hole, the location of which is now determined
to within $\pm$ 1.3 mas (10 AU).
Consequently, the dark mass at the center of the Milky Way has become
the most ironclad case of a supermassive black hole at the center of
any normal type galaxy.  

The precision of the proper motion and radial velocity measurements opens up 
additional new realms for dynamical studies in the Galactic Center.
First is the possibility of doing a full orbital
model, which also solves for the distance to the central black 
hole as  well as its motion along the line of sight (Salim \& Gould 1999; 
Ghez et al.\ 2003).  
While solving for the motion along the line of sight will require 
several more years of radial velocity data on preferably several stars, only one more year of both 
astrometric measurements and radial velocity measurements for S0-2 alone 
should provide the most direct and precise estimate of the distance to the 
Galactic Center (see, e.g., Eisenhauer et al.\ 2003).
A second opportunity is the possibility of detecting deviations from a 
Keplerian orbit.
These might arise from 
precession of the periapse distance due
to general relativistic effects (Jaroszynski 1998; Fragile \& Matthews 2000),
which would require the discovery of a star with a significantly 
smaller periapse passage than has been found so far, or, more likely, an 
extended
mass distribution (Rubilar \& Eckart 2001), in the form of either an entourage
of stellar remnants surrounding the central supermassive black hole,
a spike of dark matter particles (Gondolo \& Silk 1999; Ullio et al.\ 2001;
Gnedin \& Primack 2004) or a
binary black hole.

The stars that have been the tracers of the gravitational potential 
are themselves
quite interesting.  Their spectral features suggest that they are 
young ($<$10 Myr).  Since these stars currently reside in a region that
is inhospitable to star formation,  they are either old stars whose appearance
has been significantly altered or they are young stars formed by a
mechanism that is able to circumvent the challenges presented by the 
central black hole.  This study, for the first time, uses the 
kinematics of stars in the Sgr A* cluster to shed light on this paradox.
Among the notable properties are eccentricity, angular momentum,
and apoapse distributions that show
no statistically significant departures from an isotropic 
distribution.  This differs significantly from the He-I emission line stars, which appear to be co-planar.   It therefore appears that
the two populations of young stars in the vicinity of
the Galactic Center black hole -
the Sgr A* cluster stars and the He-I emission line stars - 
formed by different mechanisms.  In particular, it is
unlikely that the Sgr A* cluster stars formed from a disk. 
Additional orbits
will help to clarify the ensemble kinematics of this unusual group 
of stars, which reside in a particular complex region.

\acknowledgments
The authors thank Joel Aycock, Randy Campbell, Bob Goodrich, David LeMignant, 
Chuck Sorensen, and Peter Wizinowich at the Keck Observatory for
their help in obtaining the new observations,  
Gary Chanan for making it possible to phase 
the Keck Telescope on NIRC, 
Mark Reid for advance information on IRS 9 and 12N,
Mike Jura, Brad Hansen, Raoul Viollier, and Shelley Wright for helpful 
conversations, and an anonymous referee for a helpful review.
Support for this work was provided by the National Science Foundation 
grant AST-9988397 and the National Science Foundation Science
and Technology Center for Adaptive Optics, managed by the University of
California at Santa Cruz under cooperative agreement No. AST-9876783,
and the Packard Foundation.
The W.M. Keck Observatory is operated as a scientific partnership among the
California Institute of Technology, the University of California and the
National Aeronautics and Space Administration. The Observatory was made
possible by the generous financial support of the W.M. Keck Foundation.
The authors also wish to recognize and acknowledge the
very significant cultural role and reverence that the summit of Mauna
Kea has always had within the indigenous Hawaiian community.  We are
most fortunate to have the opportunity to conduct observations from
this mountain.

\appendix

\section{Source Naming}

Newly identified sources are 
named here using the convention introduced by Ghez et al.\ (1998), which
was designed to directly convey relevant information about the location 
of the source relative to the position of Sgr A$^*$.  Originally, the 
Sgr A$^*$ position given by Menten et al.\ (1997) was adopted and the 
surrounding field was divided into concentric arcsecond-wide annuli 
centered on this position.  Stars lying
within the central circle, which has a radius of 1 arcsecond, were given names
S0-1, S0-2, S0-3, etc.  Stars lying in the annulus between radii of 1 to 2
arcseconds were given the names S1-1, S1-2, and so on.  The number
immediately following "S" thus refers to the inner radius of the annulus
in which the star lies. The number following the hyphen was ordered in the 
sense of increasing distance from Sgr A* within each annulus at the time of its
naming.
In this scheme, newly identified sources are named by incrementing
the number following the hyphen within each annulus and ordered
in the sense of increasing distance from Sgr A$^*$ at the time of discovery.  
Since the original
list within 1 arcsecond ended at 15, the newly identified stars begin with 16.  
S0-16, S0-17, and S0-18 were labeled by us in a recent spectroscopic paper 
(Gezari et al.\ 2001) and S0-19 and S0-20 
were first presented at the Rees Symposium ``Making Light of Gravity," held in 
Cambridge, England (July 2002).  Due to the motions of stars, 
and refinements in the location of Sgr A*, 
the current distance 
rank does not necessarily match the one at the time of discovery.

\section{Absolute Astrometry} 

Estimates of the camera's pixel scale and orientation,
as well the position of Sgr A*, require tying the relative 
measurements to an absolute reference frame.  This was done by obtaining
multiple telescope pointings that allow the construction of mosaics 
covering the positions of Sgr A* and two SiO masers, IRS 7 and IRS 10EE
in 1998 May, 1998 Aug, 1999 May, 
1999 July, 2000 May, 2000 July, 2001 May, 2001 July.  In 1999 July, a 
somewhat larger region was covered to include the positions of two additional 
masers, IRS 9 and 12N.  
%Relative infrared positions of IRS 7 and IRS 10EE
%are obtained by the same method described in \S3.2 but using the larger mosaic 
%maps and positional uncertainties determined by requiring that linear fits
%to their proper motion velocities produces a $\chi^2_{dof}$ of 1.
By combining our infrared astrometry with radio astrometry
from Reid et al.\ (2003), we derive  
a pixel scale, 20.396 $\pm$ 0.042 mas pix$^{-1}$,
a position angle of North with respect to NIRC columns in 1999 July, 
0.80 $\pm$ 0.14 degrees, and a location of Sgr A*, which is located to within
$\pm$6.4 (E-W) mas and 14 (N-S) mas;  
Table~\ref{tab_astrom} lists the positions, with respect to Sgr A*, of 
IRS 16NW and  IRS 16C.

The infrared positional uncertainties obtained in this procedure are larger
than can be explained by uncertainties in the infrared centroids of these 
bright stars or the alignment of the map to a common epoch.
This, most likely, reflects a small residual distortion in the NIRC camera
\footnote{A known distortion in the NIRC optics is corrected for in the
individual exposures before the SAA maps are made, however any distortion 
introduced by the reimager (Matthews et al.\ 1996) has not been accounted for 
and is the likely source of additional measurement error.}.
The effects of distortion are minimized in our measurements
of the Sgr A* cluster stars by always positioning them at the center of the 
field of view and carrying out the observations 
over similar ranges of parallactic angle during every observing run
\footnote{The rotator was turned off during this experiment so the direction
of North with respect to the camera, the parallactic angle, changes throughout
the night.}.
In contrast, the masers not only occupied different camera positions, but were 
measured at different times during the night,
resulting in non-constant relative position vectors on the camera 
between each maser and Sgr A* from run to run.  This, unlike the 
measurements of the stars within the Sgr A* cluster,  maximally 
sampled the effects of distortion, which amount to a 
$\sim$0.3 pixel offset from the center of the field of view to the
edge (a 0.2\% effect).  
This distortion is what dominates our uncertainties in the inferred
infrared position of Sgr A* ($\pm$6 mas), which are nonetheless 
almost a factor of two smaller that that 
obtained by Reid et al.\ (2003, $\pm$10 mas) in their infrared reference frame.
In contrast, the distortion is not a significant effect for the relative 
stellar position measurements of 
stars in Table~\ref{tab_sum}, which have a maximum displacement of 
$\sim$0\farcs 3 over the course of this study and therefore experience 
at most a $\sim$0.6 mas offset from distortion.

\pagebreak

\pagebreak

\begin{deluxetable}{lllllllrrlrl}
\tabletypesize{\scriptsize}
\tablewidth{0pt}
\tablecaption{Summary of Sources Identified within 0\farcs 4 of the Central Dark Mass \label{tab_sum}}
\tablehead{
	\colhead{Star}&
	\colhead{Other} &
	\colhead{$<K>$ } &
	\multicolumn{2}{c}{\# Epochs} & 
	\multicolumn{4}{c}{Closest Measured Position\tablenotemark{b}} &
	\colhead{$< \sigma_{pos} >$}  &
	\colhead{$\Delta {\chi}^2~$\tablenotemark{c}}&
	\colhead{Notes}\\
	\colhead{Name} &
	\colhead{Name\tablenotemark{a}} &
	\colhead{} &
	\colhead{$N_{det}$} &
	\colhead{$N_{fit}$} &
	\colhead{Date} &
	\colhead{R} &
	\colhead{$\Delta$RA} &
	\colhead{$\Delta$Dec}  &
	\colhead{} \\ 
	\colhead{} \\ 
	\colhead{} &
	\colhead{} &
	\colhead{(mag)} &
	\colhead{} &
	\colhead{} &
	\colhead{(year)} &
	\colhead{(arcsec)} &
	\colhead{(arcsec)} &
	\colhead{(arcsec)}  &
	\colhead{(arcsec)}   &
	\colhead{} &
	\colhead{}
}
\startdata
S0-2	& S2   & 14.0 $\pm$ 0.2 & 22 & 18\tablenotemark{f} & 2000.309 & 0.012 & -0.065 &  0.059 &  0.001 & 6132 & \\
S0-16   & S14  & 15.5 $\pm$ 0.3 & 18\tablenotemark{d} & 17\tablenotemark{f} & 2000.305 & 0.006 & 0.005 & -0.001 & 0.002 & 3570 & New\tablenotemark{g}\\
S0-19	& S12  & 15.5 $\pm$ 0.2 & 13\tablenotemark{d,e} & 12\tablenotemark{f} & 1995.439 & 0.036 & 0.015 & -0.033 & 0.002 & 936 & New\tablenotemark{g}\\
S0-20   & S13  & 15.7 $\pm$ 0.2 & 15\tablenotemark{e} & 15 & 2003.682 & 0.147 & -0.136 & 0.057 & 0.004 & 286 & New\tablenotemark{g}\\
S0-1	& S1   & 14.6 $\pm$ 0.1 & 22 & 22 & 1998.505 & 0.131 & -0.117 & -0.060 & 0.001 & 281 & \\
S0-4 	& S8   & 14.4 $\pm$ 0.1 & 22 & 21\tablenotemark{f} & 1995.439 & 0.290 &  0.255 & -0.137 &  0.001 & 53  & \\
S0-5    & S9   & 15.1 $\pm$ 0.2 & 21\tablenotemark{e} & 20\tablenotemark{f} & 1995.439 & 0.316 & 0.169 & -0.267 & 0.002 & 35 & \\
S0-23	& ID7	  & 16.7 $\pm$ 0.2 & 9\tablenotemark{d,e} & 9 & 1996.485 & 0.157 & -0.024 & -0.155 & 0.005 & 14 & New\tablenotemark{g}\\
S0-25   & ID9	  & 16.4 $\pm$ 0.3 & 11\tablenotemark{e} & 11 & 1998.771 & 0.364 & 0.262 & 0.253 &  0.006 & 10 & New\tablenotemark{g} \\
S0-8    & ID14 & 15.7 $\pm$ 0.2 & 20\tablenotemark{e} & 20 & 2003.303 & 0.390 & -0.296 & 0.253 & 0.003 & 7 & New \\
S0-17   & \nodata & 15.8 $\pm$ 0.2 & 16\tablenotemark{e} & 6\tablenotemark{f} & 2003.682 & 0.115 & 0.028 & -0.112 & 0.004 & 6 & New\\
S0-26   & ID12    & 15.1 $\pm$ 0.2 & 19\tablenotemark{d,e} & 19 & 1997.367 & 0.385 & 0.366 & 0.120 & 0.002 & 6 & New\tablenotemark{g} \\
S0-22	& \nodata & 16.8 $\pm$ 0.4 & 7\tablenotemark{d,e} & 7 & 2001.572 & 0.093 & 0.031 & -0.088 & 0.01 & 5 & New \\
S0-24   & \nodata & 15.7 $\pm$ 0.2 & 5\tablenotemark{d} & 5 & 1998.505 & 0.283 & 0.244 & 0.142 &  0.008 & 2 & New \\
S0-6    & S10  & 14.2 $\pm$ 0.1 & 22 & 22 & 2003.682 & 0.378 & 0.058 & -0.374 & 0.001  & 0 & \\
S0-21 	& \nodata & 16.1 $\pm$ 0.3 & 3\tablenotemark{h} & \nodata  & 1999.560 & 0.009 & -0.007 & -0.006 & 0.006 & \nodata & New \\
S0-3	& S4   & 14.4 $\pm$ 0.2 & 22 & \nodata\tablenotemark{f} & 1995.439 & 0.180 &  0.149 &  0.101 &  0.001 & \nodata & \\ 
\enddata
\tablenotetext{a}{Other names taken from Eckart \& Genzel (1997) and
Sch\"odel et al.\ (2003).}
\tablenotetext{b}{Minimum measured projected separation (1995 - 2003)
from the dynamical center, whose location is reported in \S3.3 and Table 2.
}
\tablenotetext{c}{
$\Delta \chi^2$ is the difference between the total $\chi^2$ value
resulting from the best linear fit and the total $\chi^2$ from the
best second order polynomial fit.  Sources with $\Delta \chi^2$ greater
than 15 are considered to have significant proper motions accelerations. 
}
\tablenotetext{d}{The following stars have missing measurements due to 
stellar confusion: S0-16 (1995 June - 1997 May) due to S0-3,
S0-19 (1998 April - 1999 May) due to S0-2,  
S0-22 (2002 July - 2003 September) due to S0-17,
S0-23 (1997 May -  1998 October ) due to S0-17,
S0-24 (1998 May, 1998 August - 2003 September) due to S0-3, and 
S0-26 (1995 June - 19996 June) due to S0-7. 
} 
\tablenotetext{e}{The following stars have missing measurements that 
are likely due to insufficient map sensitivity 
(either in the main map or in at least one 
of the sub-maps; see \S2): 
S0-5 (1996 June), S0-19 (1996 June, 2000 April, 2003 April),
S0-20 (1996 June, 1998 April, May, 1999 May, 2000 July, October 2003 July),
S0-8 (1996 June \& 1998 April), S0-17 (1995 June, 1997 May, 2000 October, 
2001 July, 2002 April, May  [2001-2002 points may be missed in our 
analysis due to confusion with S0-22, causing {\it both} their correlations 
to be below our detection threshhold]), S0-22 (1995 June, 1996 June,
1997 May, 1998 July, August, October, 1999 May, 2000 October, 2001 July, 
2002 April, May [see note on S0-17]), S0-23 (1995 June, 2000 April, 2002 April,
2002 July - 2003 September), S0-25 (1995 June, 1996 June, 1998 April, July,
2000 July - 2001 May, 2002 April, 2002 July, 2003 April, September), and
S0-26 (2000 April).  
}
\tablenotetext{f}{Measurements are dropped from the
final proper motion fits as a result of 
significant astrometric biases due to nearby stars, as detailed in
table footnote {\it d} (except S0-2 in 1998 April and 1999 May)
as well as 
measurements of S0-16 in 1998 August and S0-19 in 1999 July) 
or underestimated uncertainties
(S0-4 in 1998 October and S0-5 in 2000 May).  The procedure for
identifying measurements to exclude is described in \S3.2; both 
S0-3 and S0-21 have too few ($<$3) points free from
potential biases to carry out this procedure. 
}
\tablenotetext{g}{
These stars are independently identified in Sch\"odel et al.\ (2003).
}
\tablenotetext{h}{
It is difficult to assess explicitly why other measurements of S0-21 were
not made, because of the lack of orbital information from
only three measurement and the significant stellar confusion at its
location.}
\end{deluxetable}

\begin{deluxetable}{ll}
\tablewidth{0pt}
\tablecaption{Central Dark Mass Properties from Simultaneous Orbital 
Fit to Multiple Stars\label{tab_sol_common}}
\tablehead{\colhead{Parameter}&
	   \colhead{Estimated Value}   
}
\startdata
Mass ($10^6 (\frac{R_o}{8kpc})^3 M_{\odot}$)	&  3.67 $\pm$ 0.19	\\
Position with Respect to S0-2 in 2003.0 (mas): 			&  \\
$~ ~ ~ ~ ~ ~ \Delta r_{RA}$  			&   -36.5 $\pm$ 1.6	\\
$~ ~ ~ ~ ~ ~ \Delta r_{DEC}$ 			&   -53.34 $\pm$ 0.95	\\
Proper Motion Relative to Central Cluster (mas y$^{-1}$):				& \\
$ ~ ~ ~ ~ ~ ~ V_{RA}$				&  0.87 $\pm$ 0.46 \\
$ ~ ~ ~ ~ ~ ~ V_{DEC}$			&  1.16 $\pm$ 0.57 \\
\enddata
\end{deluxetable}

\clearpage

\begin{landscape}
\begin{deluxetable}{llllllllll}
\tabletypesize{\scriptsize}
\tablewidth{0pt}
\tablecaption{Stellar Orbital Parameters from Simultaneous Orbital Fit\label{tab_sol_individ}}
\tablehead{
	   \colhead{Star}&
	   \colhead{$P$} &
	   \colhead{$A$\tablenotemark{a}} & 
           \colhead{$T_o$}  &
           \colhead{$e$}  &
	   \colhead{$i$} &
	   \colhead{$\Omega$\tablenotemark{b}} &
           \colhead{$\omega$} &
	   \colhead{$q$\tablenotemark{a}} &
	   \colhead{$Q$\tablenotemark{a}} \\
	   \colhead{}&
	   \colhead{(yrs)} &
	   \colhead{(AU)} &
           \colhead{(yrs)}  &
           \colhead{}  &
	   \colhead{(deg)} &
	   \colhead{(deg)} &
           \colhead{(deg)} &
	   \colhead{(AU)} &
	   \colhead{($\times 10^3$ AU)}  
}
\startdata
%\cutinhead{Class I orbits} 
S0-2  & 14.53  (0.65)	& 919 (23) 	& 2002.308 (0.013) & 0.8670 (0.0046) 
& 135.2 (1.2) & 44.0 (1.3) & 242.8 (2.1) & 122.2 (2.7) & 1.715 (0.047) \\
S0-16 & 36 (17) 	& 1680 (510) 	& 2000.201 (0.025) & 0.974 (0.016)   
& 102.7 (2.2) & 44.8 (1.9) & 155.9 (3.8) & 45 (16)     & 3.3 (1.0) \\
S0-19 & 37.3 (3.8)	& 1720 (110) 	& 1995.758 (0.050) & 0.833 (0.018)   
& 37.0 (4.1)  & 10.0 (8.5) & 173.8 (8.5)   & 287 (25)    & 3.15 (.22) \\
%\cutinhead{Class II orbits}
\hline
S0-20 & 43 (45)         & 1900 (1400) 	& 2005.4 (3.6)	   & 0.40 (0.21)     
& 23 (38)	& 66 (49)	& 260 (100)    & 1160 (490)  & 2.7 (2.3) \\
S0-1  & 190 (180)	& 5100 (3200) 	& 1994.04 (0.52)	& 0.70 (0.21) 
& 121.8 (1.3) & 137.4 (7.6)	& 204 (13)    & 1530 (180)  & 8.7 (6.5) \\
S0-4  & 2600 (130,000)	& 30,000 (950,000)& 1987.1 (7.5)   & 1.00 (0.15) 
& 47 (63)   & 40 (360) & 280 (300)  & 140 (270)   & 60 (1,900) \\
S0-5  & 9,900 (430,000) 	& 70,000 (2,100,000) & 2004.5 (4.7)   & 1.0 (1.3)
& 84.0 (2.2)   & 153.7 (1.1) & 356 (11) & 3206 (79) & 100 (4,200)\\
\enddata
\tablecomments{The first 3 stars represent a single joint solution.  
Each star below the line represents a four-star solution with the
first 3 stars (see \S3.3.1).  Values in parentheses are 1$\sigma$ uncertainties 
from the covariance matrix that takes into account the measurement 
uncertainties.  
%When two values are given, the uncertainties are asymmetric, with the 
%first number giving the low 1$\sigma$ value and the other the high 1$\sigma$
%value.
}
\tablenotetext{a}{The semi-major axis ($A$),
periapse distance ($q$), and
apoapse distance ($Q$) are not independent variables;
they are
reported 
here for convenience and to provide a proper accounting of 
their uncertainties.  These quantities assume a distance of 8 kpc and
their uncertainties do not include the uncertainty associated with 
this distance.} 
\tablenotetext{b}{The position angle of the nodal point is given for the 
node lying in the 
Eastern quadrants, except for S0-2 where it is the ascending node.}
\end{deluxetable}
\end{landscape}

\begin{deluxetable}{llcrrrr}
\tabletypesize{\scriptsize}
\tablewidth{0pt}
\tablecaption{Absolute Astrometry of a Few Bright Sources Relative to Sgr A*- Radio\label{tab_astrom}}
\tablehead{\colhead{Source}&
	   \colhead{$t_o$} &
	   \colhead{R} &
	   \colhead{$\Delta$RA} &
	   \colhead{$\Delta$DEC} &
	   \colhead{v$_{RA}$} &
	   \colhead{v$_{DEC}$} \\
	   \colhead{} &
	   \colhead{(year)} &
	   \colhead{(arcsec)} &
	   \colhead{(arcsec)} &
	   \colhead{(arcsec)} &
	   \colhead{(mas y$^{-1}$)} &
	   \colhead{(mas y$^{-1}$)} 
}
\startdata
IRS 16NW  & 2000.126 & 1.21 & 0.0183 $\pm$ 0.0071 & 1.212 $\pm$ 0.014 &  6.37 $\pm$ 0.18 & 1.47 $\pm$ 0.12 \\
IRS 16C   & 2000.126 & 1.23 & 1.132 $\pm$ 0.0069 & 0.484 $\pm$ 0.015 & -8.679 $\pm$ 0.095 & 7.32 $\pm$ 0.12 \\
\enddata
\end{deluxetable}


\begin{thebibliography}{}

\bibitem{} Alexander, T. \& Kumar, P. 2001, \apj, 549, 948

\bibitem{bs99} Backer, D. C., \& Sramek, R. A. 1999, \apj, 524, 805

\bibitem[Binney \& Tremaine(1987)]{1987gady.book.....B} Binney, J.~\& 
Tremaine, S.\ 1987, Princeton, NJ, Princeton University Press, 1987, 747

\bibitem{} Chakrabarty, D., \& Sarah, P. 2001, \aj, 122, 232

\bibitem[Christopher \& Scoville(2003)]{2003agnc.conf..389C} Christopher, 
M.~H.~\& Scoville, N.~Z.\ 2003, ASP Conf.~Ser.~290: Active Galactic Nuclei: 
From Central Engine to Host Galaxy, 389 


\bibitem{} Davies, M. B., Blackwell, R., Bailey, V. C., Sigurdsson, S. 1998, \mnras, 301, 745

\bibitem{diolaiti00} Diolaiti, E., Bendinelli, O.,
  Bonaccini, D., Close, L., Currie, D., \& Parmeggiani, G.\ 2000, \aaps,
  147, 335

\bibitem{bib4} Eckart, A., \& Genzel, R. 1997, \mnras, 284, 576

\bibitem{bib5} Eckart, A., Genzel, R., Ott, T., \&
Sch\"odel, R. 2002, \mnras, 331, 917

\bibitem{e99} Eckart, A., Ott, T., \& Genzel, R. 1999, \aap, 352, L22

\bibitem{} Eisenhauer, F., Sch\"odel, R., Genzel, R., Ott, T., Tecza, M., 
Abuter, R., Eckart, A., \& Alexander, T. 2003, \apj, 597, L121

\bibitem{} Ferrarese, L., \& Merritt, D. 2000, \apj, 539, L9

\bibitem{} Fragile, P. C., \& Mathews, J. 2000, \apj,
542, 328

\bibitem{} Freitag, M., \& Benz, W. 2002, \aap, 394, 345

\bibitem[Figer et al.\ 2000]{f00} Figer, D. F. et al.\ 2000, \apj, 533, L49

\bibitem{} Figer, D. F., McLean, I. S., \& Morris, M. 1999, \apj, 514, 202

\bibitem{} Figer, D. F., et al.\ 2002, \apj, 581, 258

\bibitem[Figer et al.(2003)]{2003ApJ...599.1139F} Figer, D.~F., et al.\ 
2003, \apj, 599, 1139

\bibitem{} Figer, D. 1995,  Ph.D. Thesis, University of California, Los Angeles

\bibitem{} Gebhardt, K. et al.\ 2000, \apj, 539, L13.


\bibitem{bib8} Genzel, R., Eckart, A., Ott, T., \& Eisenhauer, F. 1997, \mnras, 2
91, 219

\bibitem{} Genzel, R., Pichon, C., Eckart, A., Gerhard, O. E., Ott, T.,
2000, \mnras, 317, 348

\bibitem{} Genzel, R. et al.\ 2003a, \apj, 594, 812 

\bibitem{} Genzel, R. Sch\"odel, R., Ott, T., Eckart, A., Alexander, T., 
Lacombe, F., Rouan, D., \& Aschenbach, B. 2003b, \nat, 425, 934

\bibitem{} Genzel, R., Thatte, M., Krabbe, Kroker, H., \& Tacconi-Garman, L. E. 1996,
\apj, 472, 153

\bibitem[Gerhard (2001)]{g01} Gerhard, O. 2001, \apj, 546, L39

\bibitem{bib10} Gezari, S., Ghez, A.~M., Becklin, E.~E.,
Larkin, J., McLean, I.~S., Morris, M. 2002, \apj, 576, 790

\bibitem{} Ghez, A.~M., Duch\^ene, G., Matthews, K., Hornstein, S.~D., Tanner,
A., Larkin, J., Morris, M., Becklin, E.~E., Salim, S., Kremenek, T., Thompson,
D., Soifer, B.T., Neugebauer, G., McLean, I. 2003, \apj, 586, L127

\bibitem{bib12} Ghez, A.~M., Klein, B.~C., Morris, M.,
\& Becklin, E.~E. 1998, \apj, 509, 678

\bibitem{bib13} Ghez, A.~M., Morris, M., Becklin,
E.~E., Tanner, A., \& Kremenek, T. 2000, \nat, 407, 349

\bibitem{} Ghez, A.M., Wright, S.A., Matthews, K., Thompson, D., LeMignant, D.,
Tanner, A., Hornstein, S.D., Morris, M., Becklin, \& Soifer, B. T. 2004
\apjlett, 601, L159

\bibitem{}  Gondolo, P., \& Silk, J. 1999,
\prl, 83, 1719

\bibitem{}
Gould, A., \& Quillen, A. C. 2003, \apj , 592, 935

\bibitem[Gnedin \& Primack(2004)]{2004PhRvL..93f1302G} Gnedin, O.~Y.~\& 
Primack, J.~R.\ 2004, Physical Review Letters, 93, 061302

\bibitem{} Greenhill, L. J., Jiang, D. R., Moran, J. M., Reid, M. J., Lo, K. Y.,
Claussen, M. J. 1995, \apj, 440, 619


\bibitem{}
Haller, J. W., Rieke, M. J., Rieke, G. H., Tamblyn, P., Close, L., \& Melia,
F. 1996, \apj, 456, 194

\bibitem{} Hansen, B. M. S., \& Milosavljevi\'c, M. 2003, \apjlett, 593, L77


\bibitem{} Hornstein, S. D., Ghez, A. M., Tanner, A., Morris, M., Becklin,
E. E., Wizinowich, P, 2002, ApJ, 577, L9

\bibitem{} Hornstein, S. D., Ghez, A. M., Tanner, A., Morris, M., \&
Becklin, E. E.  2003, ``Limits on the Short Term Variability of Sagittarius A*
in the Near Infrared," Astron. Nachr., Vol. 324, No. S1, Special Supplement
"The central 300 parsecs of the Milky Way", Eds. A. Cotera, H. Falcke, T. R.
 Geballe, S. Markoff

\bibitem{} Jackson, J. M., Geis, N., Genzel, R., Harris, A. I., Madden, S.,
Poglitsch, A., Stacey, G. J., Townes, C. H. 1993, ApJ, 402, 173

%\bibitem{} Jaroszynski, M. 1998a, Acta Astronomica, 48, 413.

\bibitem{} Jaroszynski, M. 1998, Acta Astronomica, 48, 653.

\bibitem[Kim et al.\ (2004)]{kfm04} Kim, S. S., Figer, D. F., \& Morris,  M.
2004, \apjlett, 607, L123.

\bibitem[Kim \& Morris (2003)]{km02} Kim, S. S., \& Morris, M.
2003, ApJ, 597, 312

\bibitem{}
Lacy, J. H., Townes, C. H., Geballe, T. R., \& Hollenbach, D. J., 1980,
\apj, 241, 132

\bibitem{} Lee, H.M., 1996, IAU 169, 215 

\bibitem{} Levin, Y., \& Beloborodov 2003, \apj, 590, L33

\bibitem{bib19} Lo, K.~Y., Backer, D.~C.,
Ekers, R.~D., Kellermann, K.~I., Reid, M., \& Moran, J.~M.\ 1985, \nat,
315, 124

%\bibitem{} Lo, K.~Y., Claussen, M.~J. 1983, \nat, 306, 64

\bibitem{} Maoz, E. 1998, \apj, 494, L181

\bibitem{} Matthews, K., Ghez, A. M., Weinberger, A. J., and Neugebauer, G.
1996, \pasp, 108, 615

\bibitem{} Matthews,~K. and Soifer,~B.~T. 1994, Astronomy with
Infrared Arrays: The Next Generation, ed. I. McLean, Kluwer Academic
Publications (Astrophysics and Space Science, v. 190, p. 239)

\bibitem{}
McGinn, M. T., Sellgren, K., Becklin, E. E., \& Hall, D. N. B., 1989, \apj,
338, 82

%\bibitem{}McMillian, S., \& Portegies Zwart 2003, \apj, 596, 314. 

\bibitem{} Menten, K. M., Reid, M. J., Eckart, A., \& Genzel, R. 1997,
\apj, 475, L111

\bibitem{} Merritt, D., Ferrarese, L. 2001, \apj, 547, 140

\bibitem{} Miralda-Escud\'e, J., \& Gould, A. 2000, \apj, 545, 847

\bibitem[Morris (1993)]{m93} Morris, M., 1993, \apj, 408, 496

\bibitem{} Morris, M., Ghez, A. M., Becklin, E. E. 1999, Adv. Spa. Res., 23,
959


\bibitem{} Miyoshi, M., Moran, J. M., Hernstein, J., Greenhill, L., Nakai, N.,
Diamond, P., \& Inoue, M. 1995, \nat, 373, 127

\bibitem{} Munyaneza, F. \& Viollier, R. D. 2002, \apj, 564, 274.

\bibitem{} Nayaksin, S. \& Cuadra, J. 2004, \aap, submitted (astro-ph/0409541)

\bibitem{} Portegies Zwart, S. F., \& McMillian, S. L. W.  2002, \apj, 576, 899

\bibitem{} Portegies Zwart, S. F., McMillian, S. L. W., Gerhard, O. 2003, \apj, 593, 352. 

\bibitem[Rasio, Freitag, \& G{\" u}rkan(2004)]{2004cbhg.symp..138R} Rasio, 
F.~A., Freitag, M., \& G{\" u}rkan, M.~A.\ 2004, Coevolution of Black Holes 
and Galaxies, 138

\bibitem{bib23} Reid, M.~J. 1993, \araa, 31, 345

\bibitem{} Ried, M.~J. \& Brunthaler, A.
2004, \apj, accepted (astro-ph/0408107)
 
\bibitem{bib25} Reid, M.~J., Menten, K.~M., Genzel, R., Ott, T.,
Sch\"odel, R., \& Eckart, A. 2003, \apj, 587, 208

\bibitem{bib24} Reid, M.~J., Readhead, A.~C.~S., Vermeulen,
R.~C., Treuhaft, R.~N. 1999, \apj, 524, 816

\bibitem{} Rubilar,  G. F., \& Eckart, A. 2001, \aap,  2001, 372, 95

\bibitem{bib28} Salim, S., \& Gould, A. 1999, \apj, 523, 633

\bibitem[Sanders (1992)]{s92} Sanders, R. H. 1992, \nat, 359, 131

%\bibitem[Sanders (1998)]{s98} Sanders, R. H. 1998, \mnras, 294, 35

\bibitem{} Sch\"onberner, D. 1981, \aap, 103, 119

\bibitem{} Sch\"onberner, D. 1983, \apj, 272, 708

\bibitem{bib31} Sch\"odel, R. et al.\ 2002, \nat, 419, 694

\bibitem{} Sch\"odel, R., Ott, T., Genzel, R., Eckart, A., Mouawad, N., \& 
Alexander, T. 2003, \apj, 596, 1015

\bibitem{} Scoville, N. Z., Stolovy, S. R., Rieke, M., Christopher, M. H., 
Yusef-Zadeh, F. 2003, \apj, 594, 294 

\bibitem{} Sellgren, K., McGinn, M. T., Becklin, E. E., \& Hall, D. N. B. 1990,
\apj, 359, 112


\bibitem{} Spergel, D. N. et al.\ 2003, \apj, submitted

\bibitem {} Tanner, A. 2004, PhD UCLA

\bibitem{} Tremaine, S.. et al.\ 2002, \apj, 574, 740

\bibitem{} Tsiklauri, D., \& Viollier, R.~D. 1998, \apj, 500, 591

\bibitem{} Ullio, P., Zhao, H., Kamionkowski, M.
2001, \prd, 64, 1302

\bibitem{} Viollier, R. 2003, Astron. Nachr., Vol.
324, No. S1, Special Supplement "The central 300 parsecs of the Milky Way",
Eds. A. Cotera, H. Falcke, T. R.  Geballe, S. Markoff

\end{thebibliography}
\end{document}